\documentclass[showpacs,preprint,aps,epsf]{revtex4}
\usepackage{graphicx,bm,amsmath,amssymb}

\def\gz{\ifmmode{Z\hskip -4.8pt Z}
    \else{\hbox{$Z\hskip -4.8pt Z$}}\fi}

\newcommand{\be}{\begin{equation}}
\newcommand{\ee}{\end{equation}}
\newcommand{\bea}{\begin{eqnarray}}
\newcommand{\eea}{\end{eqnarray}}

\begin{document}

\title{Nonequilibrium dynamics of a singlet-triplet Anderson impurity near the quantum phase transition}
\author{P.~Roura~Bas}
\affiliation{Centro At\'{o}mico Constituyentes, Comisi\'{o}n Nacional de Energ\'{\i}a At%
\'{o}mica, 1650 San Mart\'{\i}n, Buenos Aires, Argentina}
\author{A.~A.~Aligia}
\affiliation{Centro At\'{o}mico Bariloche and Instituto Balseiro, Comisi\'{o}n Nacional
de Energ\'{\i}a At\'{o}mica, 8400 Bariloche, Argentina}
\date{\today }

\begin{abstract}
We study the singlet-triplet Anderson model (STAM) in which a configuration with a doublet is
hybridized with another containing a singlet and a triplet, as a minimal model to
describe two-level quantum dots coupled to two metallic leads in effectively
a one-channel fashion. The model has a quantum phase transition which separates regions
of a doublet and a singlet ground state. The limits of integer valence of the STAM 
(which include a model similar to the underscreened spin-1 Kondo model) are
derived and used to predict the behavior of the conductance through 
the system at both sides of the transition, 
where it jumps abruptly. 
At a special quantum critical line, 
the STAM can be mapped to an infinite-$U$ ordinary Anderson model (OAM) plus a free
spin 1/2. We use this mapping to obtain the spectral densities 
of the STAM as a function of those of the OAM at the transition.
Using the non-crossing approximation (NCA), we
calculate the spectral densities and conductance through the system as a
function of temperature and bias voltage, and determine the changes that
take place at the quantum phase transition. 
The separation of the spectral density into a singlet and a
triplet part allows us to shed light on the underlying physics and to
explain a shoulder observed recently in the zero-bias conductance as a
function of temperature in transport measurements through a 
single fullerene 
molecule [Roch N {\it et al.}, 2008 Nature \textbf{453}, 633].
The structure with three peaks observed in nonequilibrium transport 
in these experiments is also explained.
\end{abstract}

\pacs{73.63.Kv, 72.15.Qm, 75.20.Hr, 73.23.Hk}
\maketitle
\email{roura@tandar.cnea.gov.ar}

\section{Introduction}\label{intro}

For nearly four decades, the Anderson model for magnetic impurities 
has been the subject of intense
study in condensed matter physics. Its extension to the lattice 
(or even the impurity model above the so called coherence temperature)
describes among others, intermediate valence systems \cite{law,gen} and heavy fermions 
\cite{ste,hews}. A bosonic
version of it has been used to describe semiconductor microcavities with
strong light-matter interaction \cite{mis,bru}. The Kondo model is derived
through a canonical transformation as an integer valence limit of the
Anderson model \cite{sw}. The Kondo effect \ is also one of the most
relevant subjects in many-body theory \cite{hews}. A strong resurgence of
interest in these many-body phenomena takes place in recent years with
experimental results in nanoscale systems. Progress in nanotechnology has
made it possible to construct nanodevices in which the Kondo physics is
clearly displayed, for example in systems with one quantum dot (QD) 
\cite{gold,cron,wiel}, which constitute ideal systems with a single magnetic
impurity in which several parameters can be tuned. Scanning tunneling
spectroscopy has made it possible to probe the local density of states near
a single impurity and Fano antiresonances have been observed for several
magnetic impurities on metal surfaces \cite{man,naga,knorr,jan}. These
antiresonances observed in the differential conductance, are a consequence of a dip in
the spectral density of conduction states caused by the Kondo effect 
\cite{uj,pli,revi}. Furthermore, corrals built on the (111) surface of noble metals or
Cu have been used to project the spectral features of the Fano-Kondo
antiresonance to remote places \cite{man,revi}. The observed Fano line
shapes for one magnetic impurity on these surfaces have been reproduced by
many-body calculations \cite{uj,pli,revi,meri,meri2,lin,tri}. The essential physics
involved in these nanoscopic systems is well understood in terms of the
ordinary Anderson model (OAM). 
In the following, we denote by OAM the simplest version of the model,  
with infinite on-site Coulomb repulsion $U$,  in which a configuration with a
doublet is hybridized with a singlet. 
In particular, for systems with one QD
with an odd number of electrons, the conductance at zero bias is increased
below a characteristic Kondo temperature $T_{K}$ as a consequence of the
Kondo effect. This is a usual feature of single-electron transistors built
with semiconductor QD's \cite{gold,cron,wiel} or single molecules \cite{liang}.

In a QD with an even number of electrons, in many cases, the ground state is
a singlet with all dot levels either doubly occupied 
with both spin projections or empty. In this case, as a gate
voltage of either sign is applied, the system goes to a configuration with
an odd number of electrons and a doublet ground state. Therefore, the OAM
still describes the system at intermediate and even electronic occupation.
However, in other cases with an even number of electrons, due to the strong ferromagnetic
(Hund) coupling \cite{taru}, it is energetically favorable to promote one
electron of the occupied level of highest energy 
to the next unoccupied level building a triplet state. When this
triplet is well below the other states, the system can be described by the
underscreened spin-1 Kondo model, which is exactly solvable by Bethe ansatz 
\cite{s1,meh}. As a consequence, there is a partial screening of the spin 1
that explains the zero bias Kondo peak observed experimentally in this
situation \cite{sasa,schmid,kogan}. In fact, in real QD's
one expects a second screening channel to be active below a characteristic
temperature $T^{\ast }$ suppressing the conductance for bias voltage $V$ or 
temperature $T$ such that $eV,kT<kT^{\ast }$ \cite{pus,pos}. However, comparison with
experiment suggests that $T^{\ast }$ (which depends exponentially on a small
coupling constant \cite{pos}) is very small, so that one can assume an
effective one-channel model for practical purposes \cite{pos,logan}. When a
gate voltage induces a change in the occupation in such a way that the
lowest state of the isolated dot changes from a triplet to a doublet (or
conversely), the appropriate model has the form of a generalized Anderson
model which has been used to describe valence fluctuation between two
magnetic configurations \cite{maz,mag,mag2}. Its impurity version was also solved
with Bethe ansatz \cite{gen,be1,be2}. In contrast to the OAM (which has a
singlet ground state), its ground state is a doublet.

\begin{figure}[tbp]
\includegraphics[width=8cm]{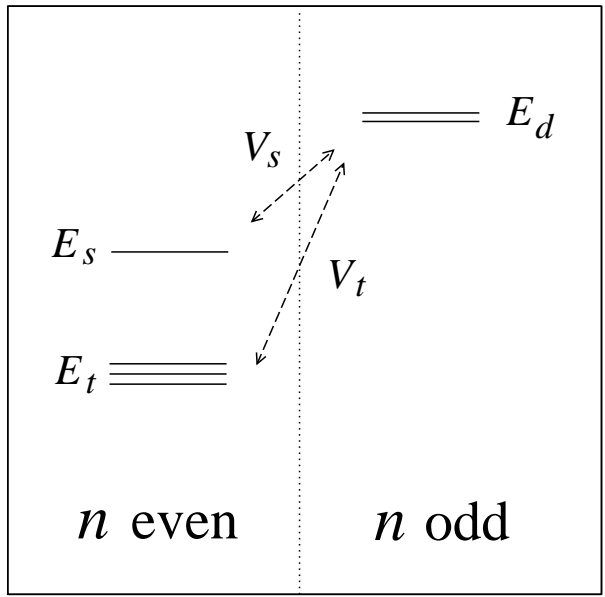}
\caption{Scheme of the lowest lying levels of the STAM: a doublet 
(a singlet and a triplet) for the configuration with an odd (even) number of particles $n$.}
\label{scheme}
\end{figure}

The physical picture becomes more complex and also more interesting when
singlet and triplet states of the configuration with even number of
electrons in the dot lie close in energy and no one of them can be
neglected (see Fig. \ref{scheme}). We call the model that describes the fluctuations between these
states and an odd-particle doublet,  the singlet-triplet Anderson model
(STAM). Again, rigorous results for this model can be borrowed from previous
studies of intermediate valence systems. Allub and Aligia proposed the model
to describe the low energy physics of Tm impurities fluctuation between the
4f$^{12}$ and 4f$^{13}$ configurations in a cubic crystal field \cite{allub}.
Using the numerical renormalization group (NRG) the authors found a
singlet or a doublet ground state depending on the parameters. Therefore,
the system has a quantum phase transition when the wave function is
forced to evolve continuously between these two competing ground states.

Quantum phase transitions is another topic of great interest in 
condensed matter physics \cite{sach}. 
Recently, Roch \textit{et al.} performed several transport
measurements through a C$_{60}$ QD with even occupancy inserted in a
nanoscale constriction \cite{roch}.
They were able to tune the parameters in
such a way that a clear manifestation of the above mentioned 
quantum phase transition 
was observed. The differential conductance $dI/dV$ as
a function of temperature and bias voltage has been measured at both
sides of the transition \cite{roch}. On the singlet side,
a dip in the conductance at $V=0$ is observed in agreement with theoretical
expectations on models similar to the STAM \cite{logan,hof,paa} as well as
non-equilibrium measurements performed in carbon nanotubes \cite{paa}. On
the other side of the transition, $dI/dV$ as a function of $V$ shows a
structure with three peaks that has not been quantitatively explained yet.
We have obtained recently a symmetric three-peak structure \cite{ro0}, but
the source of the observed asymmetry remains to be investigated. As the
temperature $T$ is decreased, the zero bias conductance $G(T)$ first
increases, then it shows a shoulder or a plateau and then increases again. The
authors state that this behavior is not understood and speculate that the
increase at the lowest temperatures might be due to the opening of 
another parallel transport mode \cite{roch}. This plateau in the conductance was presented
by us in a short paper \cite{ro0}.

Recently, a comprehensive study of the physics of a two level quantum dot,
using NRG has been presented \cite{logan}. This model contains the STAM as a
limiting case, when higher energy states can be neglected. The advantage of
the STAM is that it has fewer parameters and is the minimal model to
describe the quantum phase transition 
when charge fluctuations are allowed. An important result
of the work of Logan \textit{et al.} is the derivation of an extended
Friedel-Luttinger sum rule which relates the occupation of the dot with the
zero bias conductance at very small temperatures \cite{logan}. However, in
this work, results at finite bias were obtained using approximate
expressions and the equilibrium spectral density and (as previous works) the
results do not provide an interpretation of the above mentioned experimental
findings of Roch \textit{et al.} Because of the difficulties in extending
robust techniques to the nonequilibrium case (discussed for example in Refs. 
\cite{ro0} and \cite{none}), very few studies of this problem for finite
bias voltage exist \cite{paa,ro0}.

In this paper, we present several analytical results which shed light on the behavior
of the conductance near the quantum phase transition. 
We also present numerical results obtained 
using the non-crossing approximation (NCA), which provide an interpretation of the recent 
experiments of equilibrium and nonequilibrium conductance in C$_{60}$ QD's near the 
quantum phase transition \cite{roch}. 

In Section \ref{model} we explain the model and is application to multilevel QD's.
In Section \ref{kondo} we derive the integer valence limits of the model by means of
canonical transformations, and use known results of the ensuing effective models to predict the
behavior of the conductance at both sides of the quantum phase transition.
The self-consistent system of equations of the NCA
approximation and the expression that gives the current through a system 
described by the model, are presented in Section \ref{nca}.
In Section \ref{exact} we describe how the STAM for a particular set of quantum
critical points, can be mapped into an OAM plus a free spin, and derive useful results 
from this mapping. In particular a formula is derived, which allows 
one to calculate the spectral densities of the
STAM in terms of those of the OAM. We also show that the NCA equations satisfy 
exact results derived from this mapping. 
Section \ref{res} contains the numerical results obtained with the NCA, and comparison 
with experiment and previous works.
Section \ref{sum} is a summary and discussion.

\section{The model}\label{model}

\subsection{The mixed valence impurity}\label{tm}

As originally derived for Tm impurities in a cubic crystal field \cite{allub}, 
the STAM hybridizes the lowest states of the 4f$^{12}$ configuration, 
a $\Gamma _{1}$ singlet and $\Gamma _{4}$ triplet, with a doublet ($\Gamma _{6}$
or $\Gamma _{7}$) of the 4f$^{13}$ configuration. The fact that only two neighboring configurations
are allowed implies that infinite Coulomb repulsion $U$ is assumed.
This assumption is taken in all models discussed in this paper. 
Using the notation 
$|SM\rangle $, where $S$ is the spin and $M$ its projection, we represent the
states of the 4f$^{12}$ configuration, as $|00\rangle $ for the singlet and $%
|1M\rangle $, ($M=-1$, 0 or 1) for the triplet. The doublet is denoted by
its spin 1/2 projection $|\sigma \rangle $.

The Hamiltonian can be written in the form

\begin{equation}
H=E_{s}|00\rangle \langle 00|+E_{t}\sum_{M}|1M\rangle \langle
1M|+E_{d}\sum_{\sigma }|\sigma \rangle \langle \sigma |+H_{\text{band}}+H_{%
\text{mix}},  \label{ham}
\end{equation}%
where $H_{\text{band}}$ is a band of extended states

\begin{equation}
H_{\text{band}}=\sum_{k\sigma }\epsilon _{k}c_{k\sigma }^{\dagger
}c_{k\sigma },  \label{hband}
\end{equation}%
and $H_{\text{mix}}$ is the hybridization. In the following we assume $k$
independent matrix elements. Extensions to a more general case is
straightforward within the NCA \cite{ro1,ro2}. We can also assume full
rotational symmetry and then, the form of $H_{\text{mix}}$ is determined by
Clebsch-Gordan coefficients \cite{gen}. Calling 
$c_{\sigma }^{\dagger }=\sum_{k}c_{k\sigma }^{\dagger }/\sqrt{N}$ 
one obtains

\begin{eqnarray}
H_{\text{mix}} &=&\{[V_{s}|\uparrow \rangle \langle 00|-V_{t}(|\uparrow
\rangle \langle 10|+\sqrt{2}|\downarrow \rangle \langle 1-1|)]c_{\uparrow }
\notag \\
&+&[V_{s}| \downarrow \rangle \langle 00|+V_{t}(|\downarrow \rangle \langle
10|+\sqrt{2}|\uparrow \rangle \langle 11|)]c_{\downarrow }+\text{H.c}.\}. 
\label{hmix1}
\end{eqnarray}%
Performing an electron hole-transformation $h_{\uparrow }=-c_{\downarrow
}^{\dagger }$, $h_{\downarrow }=c_{\uparrow }^{\dagger }$, $H_{\text{mix}}$
takes the equivalent form

\begin{eqnarray}
H_{\text{mix}} &=&\{V_{s}(h_{\uparrow }^{\dagger }|\downarrow \rangle
-h_{\downarrow }^{\dagger }|\uparrow \rangle )\langle 00|+V_{t}[(h_{\uparrow
}^{\dagger }|\downarrow \rangle +h_{\downarrow }^{\dagger }|\uparrow \rangle
)\langle 10|  \notag \\
&+&\sqrt{2}(h_{\uparrow }^{\dagger }| \uparrow \rangle \langle
11|+h_{\downarrow }^{\dagger }|\downarrow \rangle \langle 1-1|)]+\text{H.c}%
.\},  \label{hmix2}
\end{eqnarray}
which is more transparent: the states $|\sigma \rangle $ may be though as having one particle
(4f hole in the Tm case) and when another particle comes from the band, a
localized two-particle singlet $|00\rangle $ or a component of the triplet 
$|1M\rangle $ is formed. In any case, it is clear that the above
transformation allows to treat with the same Hamiltonian the cases in which
the configuration with the doublet has either one more particle or one less
particle than the other one.

We can assume that $V_{s}>0$ changing if necessary, the phase of  
$|00\rangle $. Similarly we assume $V_{t}>0$. For $E_{t}\rightarrow +\infty $, 
the model reduces to the OAM. For $E_{s}\rightarrow +\infty $, the model
describes valence fluctuations between two magnetic configurations 
\cite{gen,maz}. In both limits, for constant density of conduction states, the
model is exactly solvable (by the Bethe ansatz) and the ground state is a
singlet (doublet) in the first (second) case \cite{gen,be1,be2}. Thus, the model
has a quantum phase transition as a function of $E_{s}-E_{t}$. The position of the transition
depends on the other parameters of the model, leading to a quantum critical
surface that can be determined calculating the magnetic susceptibility at 
$T\rightarrow 0$ using numerical renormalization group (NRG) \cite{allub}.
However, as shown in Section \ref{exact}, if $V_{t}=V_{s}$, the transition takes place exactly at 
$E_{s}-E_{t}=0$, independently of the value of $E_{d}$. In addition, along
this quantum critical line \cite{note}, the model can be mapped into an OAM plus a free spin 1/2.

\subsection{The multilevel dot}\label{dot}

In a dot hybridized with two leads, the triplet is formed 
from the singlet with lowest energy
by promoting an
electron from the highest occupied level, which we denote as $a$, 
to the highest unoccupied level $b$. One can restrict to these two levels. The states of these two
levels are hybridized with the bands of the two leads, left ($\nu =L$) and
right ($\nu =R$) described by

\begin{equation}
H_{\text{band}}=\sum_{\nu k\sigma }\epsilon _{\nu k}h_{\nu k\sigma
}^{\dagger }h_{\nu k\sigma },  \label{hb2}
\end{equation}%
through the following term in the Hamiltonian

\begin{equation}
H_{\text{mix}}=\sum_{\nu \sigma }\left[ (V_{\nu }^{a}a_{\sigma }^{\dagger
}+V_{\nu }^{b}b_{\sigma }^{\dagger })h_{\nu \sigma }+\text{H.c.}\right],
\label{hm2}
\end{equation}
where $h_{\nu \sigma }^{\dagger }=\sum_{k}h_{\nu k \sigma }^{\dagger }/\sqrt{N}$. 
We assume $V_{L}^{a}V_{R}^{b}=V_{L}^{b}V_{R}^{a}$, so that only one
conduction channel

\begin{equation}
h_{\sigma }=(\sum_{\nu }V_{\nu }^{\eta }h_{\nu \sigma })/[(V_{L}^{\eta
})^{2}+V_{R}^{\eta })^{2}]^{1/2}\text{ }(\eta =a\text{ or }b)  \label{cond}
\end{equation}
hybridizes with the dot states. In general, also the orthogonal linear
combination of $h_{\nu \sigma }$ plays a role and 
\textquotedblleft screens \textquotedblright\ 
the remaining doublet ground state when the
localized triplet is well below the singlet, leading to a singlet ground
state \cite{pus,pos}. However, as mentioned in Section \ref{intro}, the
characteristic energy scale involved in this second screening $T^{\ast }$
might be exponentially small. As argued before \cite{pos}, this is likely
the case of previous experiments. 
As our results will show (Section \ref{res}), the one-channel case also
describes the recent
transport measurements in C$_{60}$ QD's \cite{roch}: the theory in the
more general two-channel case \cite{pus,pos,zar} 
predicts that the zero-bias conductance $G(T)
$ should decrease at very low temperatures and $dI/dV$ should also decrease
for the smallest applied bias voltages $V$ in contrast to the observations.
This indicates that $T^{\ast }$ is smaller than the smallest temperature in
the experiments.

Assuming that the difference between the energies of the levels $b$ and $a$ is
larger than the hybridization terms, one can retain only the lowest doublet
and neglect the singlets that contain 
at least one particle in the state $b$. Then,  
performing an electron-hole transformation if necessary, the relevant
low-energy states of the dot are

\begin{eqnarray}
|\sigma \rangle &=&a_{\sigma }^{\dagger }|0\rangle \text{, }|00\rangle
=a_{\uparrow }^{\dagger }a_{\downarrow }^{\dagger }|0\rangle \text{, }%
|11\rangle =b_{\uparrow }^{\dagger }a_{\uparrow }^{\dagger }|0\rangle \text{%
, }  \notag \\
|10\rangle &=&\frac{1}{\sqrt{2}}(b_{\uparrow }^{\dagger }a_{\downarrow
}^{\dagger }+b_{\downarrow }^{\dagger }a_{\uparrow }^{\dagger })|0\rangle 
\text{, }|1-1\rangle =b_{\downarrow }^{\dagger }a_{\downarrow }^{\dagger
}|0\rangle \text{. }  \label{st}
\end{eqnarray}
The triplet states should be kept because the ferromagnetic exchange may
render them the lowest of the configuration with even number of particles
\cite{taru,sasa,schmid,kogan,roch} (see Fig. \ref{scheme}).
Restricting the action of the fermion operators to these six states one has

\begin{eqnarray}
a_{\uparrow }^{\dagger } &=&|00\rangle \langle \downarrow |\text{, }%
b_{\uparrow }^{\dagger }=|11\rangle \langle \uparrow |+|10\rangle \langle
\downarrow |/\sqrt{2},  \notag \\
a_{\downarrow }^{\dagger } &=&-|00\rangle \langle \uparrow |\text{, }%
b_{\downarrow }^{\dagger }=|1-1\rangle \langle \downarrow |+|10\rangle
\langle \uparrow |/\sqrt{2}.  \label{op}
\end{eqnarray}%
Replacing Eqs. (\ref{cond}) and (\ref{op}) in Eq. (\ref{hm2}) one obtains
Eq. (\ref{hmix2}) with

\begin{equation}
V_{s}=[(V_{L}^{a})^{2}+V_{R}^{a})^{2}]^{1/2}\text{, }%
V_{t}=[(V_{L}^{b})^{2}+(V_{R}^{b})^{2}]^{1/2}/\sqrt{2}.  \label{vsvt}
\end{equation}%
Therefore, at equilibrium, the model takes the same form as Eq. (\ref{ham}).
The values of $E_{s}$, $E_{t}$, and $E_{d}$ are easily determined from the
on-site energies and correlations at the QD, including Hund exchange 
\cite{logan}. The nonequilibrium case is discussed in Sections \ref{cur} and \ref{nca}.

\section{The integer valence limits}\label{kondo}

When $|\min (E_{s},E_{t})-E_{d}|\gg \max (V_{s},V_{t})$, the model is at the
integer valence (or ``Kondo") limit and a Hamiltonian of the exchange type can
be derived, using a canonical transformation which eliminates $H_{\text{mix}%
} $ from the Hamiltonian and originates a term quadratic in $H_{\text{mix}}$ 
\cite{sw,bat}. Two cases can be distinguished depending on which
configuration is favored.

\subsection{Odd number of electrons}\label{odd}

For $\min (E_{s},E_{t})-E_{d}\gg \max (V_{s},V_{t})$, the doublet is
favored. In this case, the canonical transformation is a particular case of
that considered in Ref. \cite{tri} for a Cr trimer on Au(111) (a doublet
ground state with virtual charge fluctuations to singlets and
triplets). Using Eqs. (\ref{ham}), (\ref{hband}), and (\ref{hmix2}), one obtains

\begin{eqnarray}
H_{\text{odd}} &=&\sum_{k\sigma }-\epsilon _{k}h_{k\sigma }^{\dagger
}h_{k\sigma }+(F_{s}-3F_{t})h_{\sigma }^{\dagger }h_{\sigma }
+J\mathbf{s\cdot S,}  \notag \\
J &=&4(F_{s}-F_{t})\text{, }F_{\eta }=\frac{V_{\eta }^{2}}{2(E_{\eta }-E_{d})%
},  \label{hodd}
\end{eqnarray}
where $\mathbf{s=}\sum_{\alpha \beta }h_{\alpha }^{\dagger }
{\boldsymbol \sigma}_{\alpha \beta }h_{\beta }/2=\sum_{\alpha \beta }
c_{\alpha }^{\dagger }{\boldsymbol \sigma}_{\alpha \beta }c_{\beta }/2$ 
is the spin of the conduction
electrons at the QD and similarly $\mathbf{S}$ is the spin of the doublet.
This is a Kondo model with potential scattering [second term of Eq. 
(\ref{hodd})]. It is known from NRG that when $J$ is positive, it is a marginally
relevant perturbation (it grows with renormalization to $J\rightarrow
+\infty $) leading to a singlet ground state \cite{allub,wil}. Instead for
negative $J$, the exchange term is marginally irrelevant ($J\rightarrow 0$)
and the ground state is a doublet. Therefore, there is a quantum phase transition 
for $J=0$, or $F_{s}=F_{t}$ in terms of the parameters of the original model.

Moreover, at both sides of the transition, the system is a Fermi liquid, but
the phase shift at the Fermi energy is $\delta =\pi /2$ for $J>0$, but $%
\delta =0$ for $J<0$. For the one-channel case that we are considering, this
means that the zero-bias conductance is maximum (vanishing) for 
$J>0$ ($J<0$) \cite{pus}. 
Therefore, there is a jump in the conductance at the quantum phase transition. 
This
is in agreement with recent NRG results and an analysis of a generalized
Friedel-Luttinger sum rule \cite{logan}.

Note that for $V_{s}=V_{t}$, the transition is exactly at $E_{s}=E_{t}$.
It has been found by NRG that this statement is valid for any occupation of
the dot and not only for integer occupation \cite{allub}. An analytic argument 
which supports this numerical result
is given in Section \ref{split}.

\subsection{ Even number of electrons}\label{even}

For $E_{d}-\min (E_{s},E_{t})\gg \max (V_{s},V_{t})$, if in addition $%
|E_{s}-E_{t}|\gg \max (V_{s},V_{t})$, the highest lying levels between the
singlet and the triplet can be neglected, and the resulting effective model
reduces to one of the cases considered in Ref. \cite{gen}. In particular if
the triplet is the lowest in energy one has the underscreened Kondo model
with singular Fermi liquid behavior \cite{meh}.

The effective model $H_{\text{even}}$ is reacher when $E_{s}$ lies near $%
E_{t}$. In the following we assume that both energies lie well below $E_{d}$. 
In order to obtain a more transparent form of $H_{\text{even}}$ we
introduce two fictitious spins 1/2, $\mathbf{S}_{1}$ and $\mathbf{S}_{2}$,
to represent the states of the configuration with even number of particles,
in terms of the states of these two spins $|\sigma _{1}\sigma _{2}\rangle $ as follows

\begin{equation}
|00\rangle =(|\uparrow \downarrow \rangle -|\downarrow \uparrow \rangle )/%
\sqrt{2},\text{ }|11\rangle =|\uparrow \uparrow \rangle ,\text{ }|10\rangle
=(|\uparrow \downarrow \rangle +|\downarrow \uparrow \rangle )/\sqrt{2},%
\text{ }|1-1\rangle =|\downarrow \downarrow \rangle .  \label{rep}
\end{equation}%
The resulting effective Hamiltonian is

\begin{eqnarray}
H_{\text{even}} &=&\sum_{k\sigma }\epsilon _{k}c_{k\sigma }^{\dagger
}c_{k\sigma }+\frac{2V_{t}^{2}}{E_{d}-E_{t}}\mathbf{s} \cdot
(\mathbf{S_{1}}+\mathbf{S_{2}})+(E_{t}-E_{s})(\mathbf{S_{1}} \cdot \mathbf{S_{2}})  \notag \\
&&+V_{s}V_{t}\left( \frac{1}{E_{d}-E_{s}}+\frac{1}{E_{d}-E_{t}}\right) 
\mathbf{s\cdot (\mathbf{\mathbf{S}}_{1}-\mathbf{\mathbf{S}}_{2})-}\frac{%
V_{t}^{2}}{E_{d}-E_{t}}n_{0}  \notag \\
&&+\left( \frac{V_{t}^{2}}{E_{d}-E_{t}}-\frac{V_{s}^{2}}{E_{d}-E_{s}}\right)
n_{0}n_{S},  \label{heven}
\end{eqnarray}%
where $n_{0}=c_{\sigma }^{\dagger }c_{\sigma }$ and $n_{S}=|00\rangle
\langle 00|=1/4-\mathbf{S_{1}} \cdot \mathbf{S_{2}}.$

For $V_s=0$ and $E_t$ well below $E_s$, $H_{\text{even}}$ reduces to the spin 1
underscreened Kondo model plus potential scattering. This model has a doublet 
ground state and singular Fermi liquid behavior \cite{meh}. 
A Hamiltonian with the first four terms (the most relevant ones) was studied
by NRG and found to have a singlet-doublet quantum phase transition 
which is in general continuous of the Kosterlitz-Thouless
type \cite{voj} , as the original model \cite{allub}. The transition is first order
only in the particular case in which the fourth term (proportional to 
$\mathbf{\mathbf{S}}_{1}-\mathbf{\mathbf{S}}_{2}$) vanishes \cite{voj}.
This implies either $V_s=0$ or $V_t=0$ in our model.
Note that for 
$V_{s}=V_{t}$ and \ $E_{s}=E_{t}$, the spin $\mathbf{\mathbf{S}}_{2}$
decouples and the spin $\mathbf{\mathbf{S}}_{1}$ has a usual Kondo
interaction with the band and is therefore screened. An analysis of the
strong coupling fixed point for the STAM \cite{allub} and for $H_{\text{even}}$ 
without the terms proportional to $n_{0}$ \cite{voj} indicates that a
generic feature of the phase with a doublet ground state (and also the
quantum critical point) is a free spin 1/2 and a Kondo screened spin 1/2 at
low temperatures. This implies a phase shift $\delta =\pi /2$ and maximum
conductance $G$, as in the phase with a singlet ground state when the
configuration with odd number of particles is favored (see previous
subsection).

Starting at the transition and decreasing $E_{s}$, the remaining spin is
also screened in a second stage and the ground state is a singlet. The full
Fermi liquid behavior is restored, and one expects $\delta =G=0$. Another
way to think on this is that the screening of the remaining spin leads to a
Fano antiresonance with a characteristic energy scale given by the second
stage Kondo effect (see Section \ref{split}).

Note that the jump in the conductance at the transition takes place
in spite of the fact that 
the transition is continuous (as discussed above) and not first order.
The impurity contribution to the magnetic susceptibility also jumps
at zero temperature at the transition \cite{allub,voj}.

These results for the conductance agree with direct calculations using NRG 
\cite{hof,logan}

\section{The non-crossing approximation (NCA)}\label{nca}

\subsection{Representation of the Hamiltonian with slave particles}\label{slave}

The NCA has been used to study the OAM out of equilibrium \cite{win}. To
extend this formalism to the STAM, we introduce auxiliary bosons, one for
the singlet state ($s$) and three for the triplets ($t_{M}$, $M=-1$, 0, 1),
and auxiliary fermions ($f_{\sigma }$) for the doublet, in analogy to the
SU(N)$\times $SU(M) generalization of the Anderson model \cite{cox,het}. In
terms of the auxiliary operators, the Hamiltonian takes the form

\begin{eqnarray}
H &=&E_{s}s^{\dagger }s+E_{t}\sum_{M}t_{M}^{\dagger }t_{M}+E_{d}\sum_{\sigma
}f_{\sigma }^{\dagger }f_{\sigma }+\sum_{\nu k\sigma }\epsilon _{\nu
k}c_{\nu k\sigma }^{\dagger }c_{\nu k\sigma }  \notag \\
&&+\sum_{\nu k\sigma }\left[ (V_{\nu }^{s}d_{s\sigma }^{\dagger }+V_{\nu
}^{t}d_{t\sigma }^{\dagger })c_{\nu k\sigma }+\text{H.c.}\right] ,
\label{hama}
\end{eqnarray}%
($\nu =L$ or $R$), where $E_{f}=E_{d}$ and

\begin{eqnarray}
d_{s\sigma }^{\dagger } &=& f_{\sigma }^{\dagger } s,  \notag \\
d_{t\uparrow }^{\dagger } &=& -(f_{\uparrow }^{\dagger } t_{0}
+\sqrt{2}f_{\downarrow }^{\dagger } t_{-1})/\sqrt{3},  \notag \\
d_{t\downarrow }^{\dagger } &=& (f_{\downarrow }^{\dagger } t_{0}
+\sqrt{2}f_{\uparrow }^{\dagger } t_{1})/\sqrt{3},  \label{d}
\end{eqnarray}
with the constraint

\begin{equation}
s^{\dagger }s+\sum_{M}t_{M}^{\dagger }t_{M}+\sum_{\sigma }f_{\sigma
}^{\dagger }f_{\sigma }=1.  \label{con}
\end{equation}

The factor $1/\sqrt{3}$ in Eqs. (\ref{d}) was chosen to give a more symmetric form
for the NCA equations, in particular near the quantum critical line for
which the system is exactly solvable (see Section \ref{enca}). 
Comparing with Eqs.
(\ref{op}), one can realize that $d_{s\uparrow }^{\dagger }=-a_{\downarrow }$, 
$d_{s\downarrow }^{\dagger }=a_{\uparrow }$, $d_{t\uparrow }^{\dagger }=-%
\sqrt{2/3}b_{\downarrow }$, $d_{t\downarrow }^{\dagger }=\sqrt{2/3}%
b_{\uparrow }$. The representation of Eq. (\ref{hama}) was chosen in such a
way that if the triplet can be neglected (because either $E_{t}\rightarrow
+\infty $ or $V_{L}^{t}=V_{R}^{t}=0$), the model reduces to the OAM. 

As in Section \ref{dot}, 
the mixing part of Eq. (\ref{hama}) can be put in the form of Eq. (\ref{hmix1})  with $V_{s}=[(V_{L}^{s})^{2}+V_{R}^{s})^{2}]^{1/2}$, 
$V_{t}=[(V_{L}^{t})^{2}+(V_{R}^{t})^{2}]^{1/2}/\sqrt{3}.$

\subsection{Equation for the current}\label{cur}

For the calculation of the current, 
we consider a multilevel QD with at most six relevant states
of the Hilbert space, as described in Section \ref{dot}  [see Eq. (\ref{st})]. For
the case of proportionate couplings of the relevant two levels 
($V_{L}^{a}V_{R}^{b}=V_{L}^{b}V_{R}^{a}$ as we are assuming), Meir and
Wingreen \cite{meir} provided an expression for the current in a
non-equilibrium situation [Eq. (9) of Ref. \cite{meir}], which is given by a
trace of $\mathbf{\Gamma G}^{r}$, where $\mathbf{\Gamma }$ is given in terms
of $V_{\nu }^{\eta }$ and $\mathbf{G}^{r}$ is a matrix of retarded Green's
functions. In \ our case, $\mathbf{\Gamma }$ and $\mathbf{G}^{r}$ are 
4$\times $4 matrices in spin and level ($a$ or $b$) indices. 
It is easy to see that the product $\mathbf{\Gamma G}^{r}$ is the same (as it should be)
in the representation of slave particles used above [Eqs. (\ref{d})], in which the 
the index $\eta $ refers to $s$ and $t$ instead of $a$ and 
$b$ (the normalization factors in $\mathbf{\Gamma }$ and $\mathbf{G}^{r}$ cancel).

$\mathbf{\Gamma }$
is diagonal in spin index. Within the NCA, the expectation values entering 
the Green's functions decouple into fermion and boson parts \cite{win} and
both are diagonal as a consequence of SU(2) invariance of the
Hamiltonian. This means that $\mathbf{G}^{r}$ is
diagonal in level index. This allows to simplify the resulting expression, which takes
the form

\begin{equation}
I=\frac{A\pi e}{h}\int d\omega \sum_{\eta }\Gamma ^{\eta }\rho _{d}^{\eta
}(\omega )[f_{L}(\omega )-f_{R}(\omega )],  \label{i}
\end{equation}
where $\rho _{d}^{\eta }(\omega )=-\text{Im}G_{d \eta \sigma }^{r}(\omega )/\pi $ 
is the spectral density of $d_{\eta \sigma }^{\dagger }$,

\begin{equation}
\Gamma ^{\eta }=\Gamma _{R}^{\eta }+\Gamma _{L}^{\eta }\text{, with }\Gamma
_{\nu }^{\eta }=2\pi \sum_{k}|V_{\nu }^{\eta }|^{2}\delta (\omega -\epsilon
_{k})  \label{gam}
\end{equation}
assumed independent of $\omega $ within a bandwidth $D$ and zero elsewhere, 
\be
A=4\Gamma _{R}^{\eta }\Gamma _{L}^{\eta }/(\Gamma _{R}^{\eta }+\Gamma
_{L}^{\eta })^{2}\leq 1 \label{a}
\ee
(independent of $\eta $) is a parameter that
characterizes the asymmetry between left and right leads, and $f_{\nu
}(\omega )$ is the Fermi function with the chemical potential $\mu _{\nu }$
of the corresponding lead.

\subsection{Spectral densities and Green's functions}\label{sdgf}

The spectral densities of the operators $d_{\eta \sigma }^{\dagger }$
defined by Eqs. (\ref{d}) for given spin, $\rho _{d}^{s}(\omega )$ and $\rho
_{d}^{t}(\omega )$, are determined by convolutions from those of the
auxiliary particles $\rho _{\lambda }(\omega )$ with 
the lesser Green's functions $G_{\lambda}^{<}(\omega )$ 
($\lambda =s$, $t$, or $f$) as follows

\begin{equation}
\rho _{d}^{\eta }(\omega )=\frac{1}{Z}\int d\omega ^{\prime }\left(
G_{\eta }^{<}(\omega ^{\prime })\rho _{f}(\omega ^{\prime }+\omega
)+G_{f}^{<}(\omega ^{\prime }+\omega )\rho _{\eta }(\omega ^{\prime
})\right) ,  \label{rho}
\end{equation}

\begin{equation}
Z=\int d\omega \left( G_{s}^{<}(\omega )+2G_{f}^{<}(\omega
)+3G_{t}^{<}(\omega )\right) ,  \label{z}
\end{equation}%
where we define $G_{t}^{<}(\omega )$ as the
Fourier transform of $\langle t_{M}^{\dagger }(0)t_{M}(t)\rangle $ (the
result is independent of $M$ because of SU(2) symmetry) and similarly for $%
G_{s}^{<}(\omega )$ and $G_{f}^{<}(\omega )$. The spectral densities of the
auxiliary particles are given by the imaginary part of the corresponding
retarded Green's function as usual: $\rho _{\lambda }(\omega )=
-\text{Im}G_{\lambda }^{r}(\omega )/\pi $. In turn, these Green's functions

\begin{equation}
G_{\lambda }^{r}(\omega )=\frac{1}{\omega -E_{\lambda }-\Sigma _{\lambda
}^{r}(\omega )},  \label{g}
\end{equation}%
are given in terms of retarded self-energies $\Sigma _{\lambda }^{r}(\omega )$, 
which as in the case of the OAM \cite{win}, should be 
determined selfconsistently.
The imaginary parts are given by the following set of integral
equations

\begin{eqnarray}
\text{Im}\Sigma _{s}^{r}(\omega ) &=&-\int d\omega ^{\prime }~\Gamma
^{s}\rho _{f}(\omega ^{\prime })\tilde{f}(\omega ^{\prime }-\omega ),  \notag
\\
\text{Im}\Sigma _{t}^{r}(\omega ) &=&-\frac{1}{3}\int d\omega ^{\prime
}~\Gamma ^{t}\rho _{f}(\omega ^{\prime })\tilde{f}(\omega ^{\prime }-\omega
),  \notag \\
\text{Im}\Sigma _{f}^{r}(\omega ) &=&-\frac{1}{2}\int d\omega ^{\prime
}~[\Gamma ^{s}\rho _{s}(\omega ^{\prime })+\Gamma ^{t}\rho _{t}(\omega
^{\prime })]\tilde{h}(\omega - \omega^{\prime }),  \label{sigr}
\end{eqnarray}%
where

\begin{eqnarray}
\tilde{f}(\omega ) &=&[\Gamma _{L}^{\eta }f_{L}(\omega )+\Gamma _{R}^{\eta
}f_{R}(\omega )]/\Gamma ^{\eta },  \notag \\
\tilde{h}(\omega ) &=&[\Gamma _{L}^{\eta }(1-f_{L}(\omega ))+\Gamma
_{R}^{\eta }(1-f_{R}(\omega ))]/\Gamma ^{\eta },  \label{avf}
\end{eqnarray}
and $\Gamma _{\nu}^{\eta }$, $\Gamma^{\eta }$ ($\eta = s$ or $t$) are given by Eqs. (\ref{gam}).
The real part of the self energies are obtained using Kramers-Kronig
relations

\begin{equation}
\text{Re}\Sigma _{\lambda }^{r}(\omega )=\frac{1}{\pi }\mathcal{P}\int
d\omega ^{\prime }~\frac{\text{Im}\Sigma _{\lambda }^{r}(\omega ^{\prime })}{%
\omega ^{\prime }-\omega }.  \label{kk}
\end{equation}

Once the retarded self-energies are obtained solving the above system of
equations, the lesser Green's functions come from the solution of the
following integral equations

\begin{equation}
G_{d}^{<}(\omega )=|G_{d}^{R}(\omega )|^{2}\Sigma _{d}^{<}(\omega ),
\label{gel}
\end{equation}

\begin{eqnarray}
\Sigma _{s}^{<}(\omega ) &=&\frac{1}{\pi }\int d\omega ^{\prime }~\Gamma
^{s}G_{f}^{<}(\omega ^{\prime })\tilde{h}(\omega ^{\prime }-\omega ),  \notag
\\
\Sigma _{t}^{<}(\omega ) &=&\frac{1}{3\pi }\int d\omega ^{\prime }~\Gamma
^{t}G_{f}^{<}(\omega ^{\prime })\tilde{h}(\omega ^{\prime }-\omega ),  \notag
\\
\Sigma _{f}^{<}(\omega ) &=&\frac{1}{2\pi }\int d\omega ^{\prime }~[\Gamma
^{s}G_{s}^{<}(\omega ^{\prime })+\Gamma ^{t}G_{t}^{<}(\omega ^{\prime })]%
\tilde{f}(\omega-\omega^{\prime } ).  \label{sigl}
\end{eqnarray}

\subsection{Numerical details}\label{nd}

In the numerical procedure to solve the NCA equations, we have used a set of self-adjusting meshes (rather than the fixed one used in Ref. \cite{het}) to describe the spectral densities and lesser Green's functions  
of the auxiliary particles: we have evaluated $\rho _{\lambda }(\omega )$ and $G_{\lambda}^{<}(\omega )$ in the corresponding logarithmic array of discrete frequencies $\omega_{\lambda}$, built in each iteration in order to have a larger density of points near the corresponding peaks or singularities of these functions. The procedure guarantees the resolution of the sets of integral equations (\ref{sigr}) and (\ref{sigl}) to a high degree of accuracy. To calculate the spectral densities, $\rho_{d}^{s}(\omega)$ and $\rho_{d}^{t}(\omega)$, we have used two different logarithmic meshes centered at the peaks of the functions entering Eqs. (\ref{rho}).
This scheme of numerical resolution allows us to obtain the conductance at both sides of the transition, within equilibrium and non equilibrium, and for all values of the parameters considered.

The logarithmic discretization is similar to that used in NRG calculations, where only one mesh centered at the
Fermi energy $\mu_L=\mu_R$ and with an arbitrarily large number of frequencies near this energy is used \cite{bulla}.

\section{The exactly solvable case}\label{exact}

For $V_{s}=V_{t}$ and \ $E_{s}=E_{t}$, the model given by Eqs. (\ref{ham}), 
(\ref{hband}), and (\ref{hmix1}) or (\ref{hmix2}) has additional symmetries
and is exactly solvable \cite{allub}. It has been shown numerically that
these equations define a quantum critical line (a point for each
value of $E_{d}/V_{s}$) \cite{note} that separates regions of singlet and
double ground states \cite{allub}. In this Section we provide simple
analytical arguments to demonstrate these results, map the corresponding
spectral densities, and show that the NCA is consistent with these results.

\subsection{The STAM on the quantum critical line}\label{qcl}

In analogy to Eq. (\ref{rep}), 
let us consider a (probably fictitious) system, like the two-level one
considered in Section \ref{dot}, but in which the relevant singlet is $|00\rangle
=(1/\sqrt{2})(b_{\uparrow }^{\dagger }a_{\downarrow }^{\dagger
}-b_{\downarrow }^{\dagger }a_{\uparrow }^{\dagger })|0\rangle $ [instead of 
$a_{\uparrow }^{\dagger }a_{\downarrow }^{\dagger }|0\rangle $, see Eqs. 
(\ref{st})], and with mixing term 

\begin{equation}
H_{\text{mix}}=\sum_{\nu \sigma }\left[ V_{\nu }^{b}b_{\sigma }^{\dagger
}h_{\nu \sigma }+\text{H.c.}\right],  \label{hme}
\end{equation}
which does not involve the $a_{\sigma}^{\dagger }$ and $a_{\sigma }$ operators. 
Clearly, the resulting model 
[Eq. (\ref{oamd}) with $E_t=E_s$] 
is the OAM with infinite Coulomb repulsion $U$ for the
level $b$ (which is exactly solvable by Bethe ansatz \cite{gen}), while
level $a$ within the relevant Hilbert subspace [Eqs. (\ref{st}) with 
$|00\rangle $ replaced as above] reduces to a decoupled spin 1/2: $|\sigma
\rangle =a_{\sigma }^{\dagger }|0\rangle $. Within this subspace 

\begin{eqnarray}
b_{\uparrow }^{\dagger } &=&|11\rangle \langle \uparrow |+(|10\rangle
\langle \downarrow |+|00\rangle \langle \downarrow |)/\sqrt{2},  \notag \\
b_{\downarrow }^{\dagger } &=&|1-1\rangle \langle \downarrow |+(|10\rangle
\langle \uparrow |-|00\rangle \langle \uparrow |)/\sqrt{2}.  \label{ope}
\end{eqnarray}
Replacing these equations in Eq. (\ref{hme}) one obtains Eq. (\ref{hmix2})
with

\begin{equation}
V_{s}=V_{t}=[(V_{L}^{b})^{2}+(V_{R}^{b})^{2}]^{1/2}/\sqrt{2}.
\label{vsvt2}
\end{equation}
This shows the equivalence of the STAM for $V_{s}=V_{t}$ and \ $E_{s}=E_{t}$
with an OAM plus a free doublet.

\subsection{Effect of singlet-triplet splitting}\label{split}

Proceeding as above, it is easy to see that for 
$V_{s}=V_{t}=V_{\text{OAM}}/\sqrt{2}$, 
but arbitrary $E_{s}$ and $E_{t}$ the STAM given by Eqs. (\ref{ham}), (%
\ref{hband}), and  (\ref{hmix2}), except for an irrelevant constant is
mapped onto

\begin{eqnarray}
H' &=&\sum_{k\sigma }-\epsilon _{k}h_{k\sigma }^{\dagger }h_{k\sigma
}+(E_{t}-E_{d})\sum_{\sigma }b_{\sigma }^{\dagger }b_{\sigma }
+\sum_{\sigma }\left[ V_{\text{OAM}}b_{\sigma }^{\dagger }h_{\sigma }+\text{H.c.}\right] 
\notag \\
&+&Ub_{\uparrow }^{\dagger }b_{\uparrow }b_{\downarrow }^{\dagger
}b_{\downarrow }+(E_{t}-E_{s})(\mathbf{S}_{a}\mathbf{\cdot S}_{b}-1/4),
\label{oamd}
\end{eqnarray}
where $U\rightarrow +\infty $, $\mathbf{S}_{a}$ is the spin operator of the
spin 1/2 which is free for $E_t=E_s$, and similarly $\mathbf{S}_{b}=\sum_{\alpha \beta }b_{\alpha
}^{\dagger }{\boldsymbol \sigma}_{\alpha \beta }b_{\beta }/2.$

For $E_{t}=E_{s}$, clearly $\mathbf{S}_{b}$ is screened as usual in the OAM
and the ground state is a doublet which is the direct product of a Fermi
liquid singlet times the spin state $|\sigma \rangle $. If the state $b$ is
hybridized with two conducting leads (as above), the physics of the OAM
determines that the conductance $G=G_{0}\sin ^{2}\delta $, with $%
G_{0}=2e^{2}A/h$ and $\delta =\pi \langle \sum_{\sigma }b_{\sigma }^{\dagger
}b_{\sigma }\rangle /2$ \cite{pro}. In particular when the total number of
particles $\langle \sum_{\sigma }b_{\sigma }^{\dagger }b_{\sigma }\rangle
+1=2$, one has maximum conductance, in agreement with the result discussed
in Section \ref{even}.

When the last term of Eq. (\ref{oamd}) is added, one can think 
$\mathbf{S}_{b}$ as representing the spin of itinerant electrons in an effective heavy
mass Fermi liquid at low energies. In fact, comparison of a mean-field
slave-boson treatment with NRG calculations \cite{cor} show that this
picture is qualitatively correct \cite{note2} for $\langle \sum_{\sigma }b_{\sigma
}^{\dagger }b_{\sigma }\rangle +1\simeq 2$. Then, as discussed in Section
\ref{kondo}, from the physics of the ensuing effective Kondo model \cite{wil,cor} ,
when $E_{t}<E_{s}$ (ferromagnetic coupling) the exchange interaction
renormalizes to zero and the ground state continues to be a doublet, while
for $E_{t}>E_{s}$ a second screening takes place and the ground state is a
singlet. This results agrees with previous NRG results \cite{allub} and
confirms that $V_{s}=V_{t}$ and \ $E_{s}=E_{t}$ corresponds to the quantum critical line. 

If the OAM for $E_{t}=E_{s}$ is in the Kondo regime ($\langle \sum_{\sigma
}b_{\sigma }^{\dagger }b_{\sigma }\rangle \simeq 1$), one expects that
addition of a positive exchange ($E_{t}>E_{s}$) induces a Fano-Kondo
antiresonance, depressing the conductance at low temperatures \cite{pro,cor},
while nothing dramatic happens for $E_{t}<E_{s}$. This again agrees with
the results of Section \ref{kondo}. 

While as discussed above, the ground state is a doublet for $E_{t}<E_{s}$, 
we remind the reader that for realistic two-level systems at low enough 
temperatures ($T < T^{\ast }$, see Section \ref{dot}) a second screening channel
should become active leading to a screening of the remaining doublet 
and a decrease in the conductance \cite{pus,pos,zar}. 

\subsection{Mapping of the spectral densities}\label{dens}

Since the OAM out of equilibrium has been studied before \cite{none,win,hbo,rpt,mr}, 
results for the conductance of the OAM can be extended to the STAM on 
the quantum critical line
if one knows how to express the spectral densities $\rho _{d}^{s}(\omega )$
and $\rho _{d}^{t}(\omega )$ of the operators $d_{s\sigma }^{\dagger }$, 
$d_{t\sigma }^{\dagger }$ of the STAM [see Eqs (\ref{d})], which enter the
equation for the current (\ref{i}), in terms of the spectral density 
$\rho_{b}(\omega )$ of the operator $b_{\sigma }^{\dagger }$ of the OAM [included
in Eq. (\ref{oamd}) for $E_{t}=E_{s}$]. These densities are proportional to
the imaginary part of the corresponding retarded Green's functions. For
example $\rho _{b}(\omega )=-\text{Im}G_{b}^{r}(\omega )$, where $%
G_{b}^{r}(\omega )$ is the Fourier transform of $G_{b}^{r}(t)=-i\theta
(t)\langle b_{\sigma }(t)b_{\sigma }^{\dagger }+b_{\sigma }^{\dagger
}b_{\sigma }(t)\rangle $, where $\theta (t)$ is the step function. The
operators $d_{s\sigma }^{\dagger }$, $d_{t\sigma }^{\dagger }$ of the STAM
can be expressed in terms of those of the OAM using Eqs (\ref{d}) and the
mapping of operators explained in Section \ref{qcl}. 
For example

\begin{eqnarray}
d_{s\downarrow } &=&\frac{1}{\sqrt{2}}(b_{\uparrow }^{\dagger }a_{\downarrow
}^{\dagger }-b_{\downarrow }^{\dagger }a_{\uparrow }^{\dagger
})a_{\downarrow },  \notag \\
d_{t\downarrow } &=&\frac{1}{\sqrt{6}}[2b_{\uparrow }^{\dagger }a_{\uparrow
}^{\dagger }a_{\uparrow }+(b_{\uparrow }^{\dagger }a_{\downarrow }^{\dagger
}+b_{\downarrow }^{\dagger }a_{\uparrow }^{\dagger })a_{\downarrow }].
\label{map}
\end{eqnarray}%
Replacing this into expectation values like $\langle d_{\eta \downarrow
}(t)d_{\eta \downarrow }^{\dagger }\rangle $, one obtains expectation values
involving six fermion operators. Four of them correspond to $a_{\sigma
}^{\dagger }$ and $a_{\sigma }$ operators, which have no dynamics (are time
independent) and are decoupled from the remaining $b_{\sigma }^{\dagger }$
and $b_{\sigma }$ operators. Evaluating the expectation values involving $%
a_{\sigma }^{\dagger }$ and $a_{\sigma }$ operators, using spin
conservation, $\langle a_{\uparrow }^{\dagger }a_{\uparrow }a_{\downarrow
}^{\dagger }a_{\downarrow }\rangle =0$, and  assuming the paramagnetic phase
(in particular  $\langle a_{\sigma }^{\dagger }a_{\sigma }\rangle =1/2$), we
obtain after some algebra

\begin{equation}
\langle d_{\eta \downarrow }(t)d_{\eta \downarrow }^{\dagger }\rangle =\frac{%
1}{2}\langle b_{\sigma }^{\dagger }b_{\sigma }(-t)\rangle ,  \label{exp}
\end{equation}%
independently of $\eta =s$ or $t$. Proceeding in the same way for the
remaining expectation values we finally obtain

\begin{equation}
\rho _{d}^{s}(\omega )=\rho _{d}^{t}(\omega )=\frac{1}{2}\rho _{b}(-\omega ),
\label{rero}
\end{equation}%
which relates the spectral densities of the STAM to that of the OAM on the
quantum critical line.

\subsection{Mapping of the NCA equations}\label{enca}

It is interesting to note that the NCA approach for the STAM described in
Section \ref{nca}, and the corresponding one for the OAM, although they
seem to be quite different at first glance, can be related and satisfy 
Eqs. (\ref{rero}). The NCA for the OAM \cite{win} makes use of an auxiliary boson 
$\tilde{b}$ and auxiliary fermions $\tilde{f}_{\sigma }$, describing the
electron operator as $b_{\sigma }=\tilde{b}^{\dagger }\tilde{f}_{\sigma }$.
In analogy to Eqs. (\ref{gam}), the coupling to the right and left leads are
described by coupling constants $\Gamma _{R}$ and $\Gamma _{L}$. We define
the total coupling of the OAM as $\Gamma =\Gamma _{R}+\Gamma _{L}$. The
mapping between both models described above for $V_{s}=V_{t}$ and \ $E_{s}=E_{t}$
implies that $\Gamma _{\nu }^{s}=1/2\Gamma _{\nu }$, and $\Gamma _{\nu
}^{t}=3/2\Gamma _{\nu }$. Therefore, $\Gamma ^{t}=3\Gamma ^{s}=3/2\Gamma $.
As a consequence of these relations, the equations for the auxiliary bosons $s$ 
and those for $t_{M}$ have the same form 
[see Eqs. (\ref{sigr}) and (\ref{sigl})]. Moreover
the resulting equations for the self-energies take the same form as those of
the auxiliary fermions $\tilde{f}_{\sigma }$ of the OAM [Eq. (23) of Ref. 
\cite{win}], but with $f_{\nu }(\omega )$ replaced by $1-f_{\nu }(\omega )$.
Similarly, the self-energies of the auxiliary fermions $f_{\sigma }$ of the
STAM take the same form as those of the auxiliary boson $\tilde{b}$ of the
OAM, with the same change as above in the Fermi functions $f_{\nu }$. 
This implies the following relations for
the auxiliary particles

\begin{eqnarray}
\rho _{s}(\omega ) &=&\rho _{t}(\omega )=\rho _{\tilde{f}}(-\omega )\text{, }%
\rho _{f}(\omega )=\rho _{\tilde{b}}(-\omega )\text{,}  \notag \\
G_{s}^{<}(\omega ) &=&G_{t}^{<}(\omega )=G_{\tilde{f}}^{<}(-\omega )\text{, }%
G_{f}^{<}(\omega )=G_{\tilde{b}}^{<}(-\omega )\text{.}  \label{ros}
\end{eqnarray}%
The density $\rho _{b}(\omega )$ of the real fermion $b_{\sigma }$ of the
OAM is given by a convolution [Eq. (21) of Ref. \cite{win}] similar to that
defining  $\rho _{d}^{s}(\omega )$ and $\rho _{d}^{t}(\omega )$ [Eq. (\ref%
{rho})] with the replacements indicated by Eqs. (\ref{ros}), but with the
difference that $Z$ in the denominator is replaced by $Z_{\text{OAM}}=\int
d\omega \left( G_{\tilde{b}}^{<}(\omega )+2G_{\tilde{f}}^{<}(\omega )\right) 
$. Using Eqs. (\ref{z}) and (\ref{ros}), it is easy to see that $Z=2Z_{\text{%
OAM}}$. This lead to Eqs. (\ref{rero}) for the relation between spectral
densities. We have checked it numerically by an independent solution of the
NCA equations for both models.  

\section{Numerical results}\label{res}

For the numerical solution of the NCA equations, we take $V_{s}=V_{t}$, so
that independently of $E_{d}$, the quantum transition occurs at the exactly
solvable case $E_{s}=E_{t}$ described above. We take $\Gamma $, the coupling
of the OAM involved in the mapping described in the previous section, as the
unit of energy. Therefore $\Gamma ^{s}=1/2\Gamma $, and $\Gamma
^{t}=3/2\Gamma $. We take the band width of the conduction bands $D=10\Gamma 
$. At equilibrium $\mu _{L}=\mu _{R}$ and the properties of the model depend
on $E_{\eta }+\mu _{L}-E_{d}$ ($\eta =s$ or $t$) and not separately on the
individual parameters. Therefore, without loss of generality, we take 
$E_{d}=\mu _{L}=\mu _{R}=0$. Out of equilibrium, unless otherwise stated, we
assume 
$\Gamma _{R}^{\eta }=\Gamma _{L}^{\eta }=\Gamma ^{\eta }/2$ and 
$\mu _{L}+\mu _{R}=0$ (as expected for equal couplings to left and right leads).

The choice $V_{s}=V_{t}$ leaves four free parameters:
temperature $T$, bias voltage $V$ which determines the difference in
chemical potentials $\mu _{L}-\mu_{R}=eV$, $E_{t}/\Gamma $ (or $E_{s}/\Gamma $) 
which controls the valence and can be modified by a gate voltage,  and
finally $(E_{s}-E_{t})/\Gamma $ which controls the distance to the quantum critical line 
\cite{note}. 
This parameter has also been controlled experimentally by Roch {\it et al.}  \cite{roch}.
A great advantage of taking $V_{s}=V_{t}$ is that we know
exactly where the quantum transition is, while for other ratios $V_{s}/V_{t}$, 
the position of the transition has to be determined numerically 
\cite{allub}. This is very time consuming within the NCA because it is required
to solve the structure of the spectral densities at low temperatures many
times near the transition.

In this paper, we restrict our study to $E_{s}, E_{t} < E_{d}$. This means that 
the configuration with even number of particles is favored. This situation corresponds
to the most novel experimental results, in particular those of Roch {\it et al.} 
for the conductance through C$_{60}$ QD's near the quantum phase transition \cite{roch}.
In the following and for the sake of brevity we call the ``singlet side'' of the
transition the region of parameters with a singlet ground state ($E_{s}< E_{t}$ 
in our case with $V_{s}=V_{t}$), and (to be consistent with Roch {\it et al.})
we denote by  ``triplet side'' the region $E_{s} > E_{t}$ although as explained
above, the spin 1 is partially screened and the ground state is a doublet.

\subsection{The spectral densities}\label{spec} 

\begin{figure}[tbp]
\includegraphics[width=8cm]{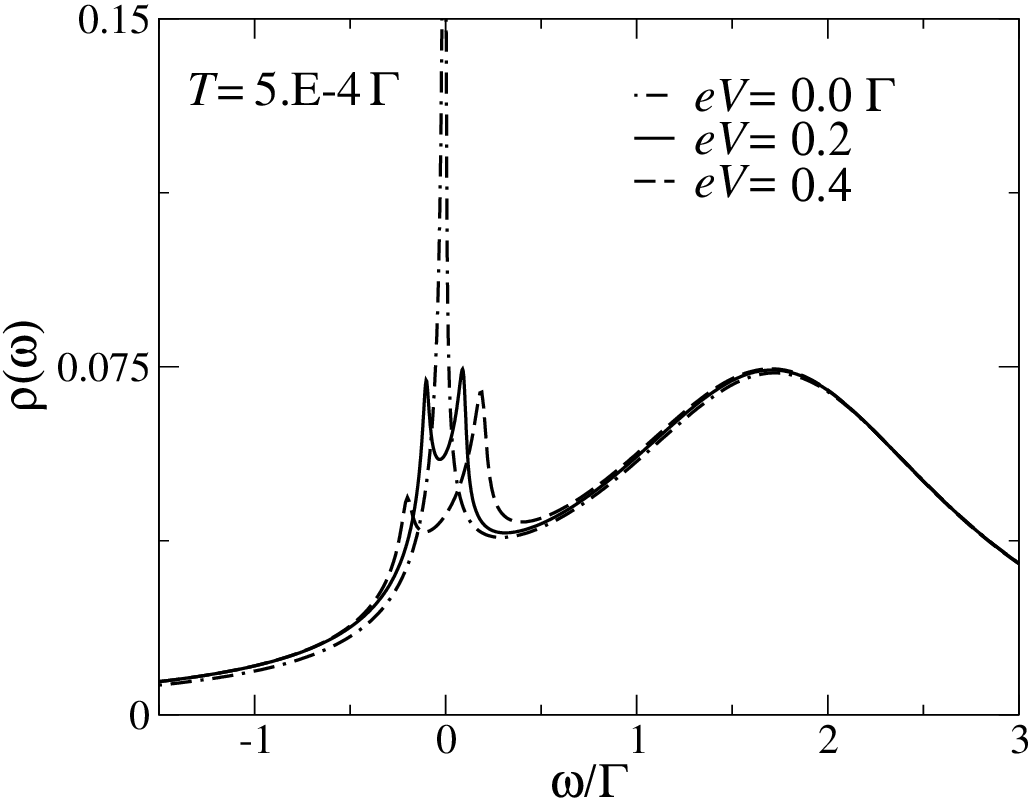}
\caption{Spectral densities as a function of frequency for $T=5 \times 10^{-4}$, 
$E_t=E_s=-2$, and three bias voltages $V$.}
\label{doam}
\end{figure}

As shown in Section \ref{cur}, the current is proportional to the 
integral of the following weighted average spectral density
\be
\rho _{d}^\text{av}(\omega )=\frac{\sum_{\eta }\Gamma ^{\eta }\rho _{d}^{\eta}(\omega )}{\Gamma ^{s}+\Gamma ^{t}}.  \label{rhoav}
\ee
However, as we will show, a study of the singlet $\rho _{d}^{s}(\omega )$ 
and triplet $\rho _{d}^{t}(\omega )$ parts of 
this average density contributes significantly to the understanding
of the numerical results.
As expected from Section \ref{enca}, within our numerical accuracy, 
the three densities, shown in Fig. \ref{doam}, coincide with the specular image
of that of the OAM times a factor 1/2 (Fig. 5 of Ref. \cite{win}).
At equilibrium, they show a peak at the Fermi level. The width of this peak allows to define a
Kondo temperature $T_{K}$. Under an applied bias voltage, the peak splits in two
near the corresponding $\mu_\nu$ as in the OAM \cite{win}.

\begin{figure}[tbp]
\includegraphics[width=12cm]{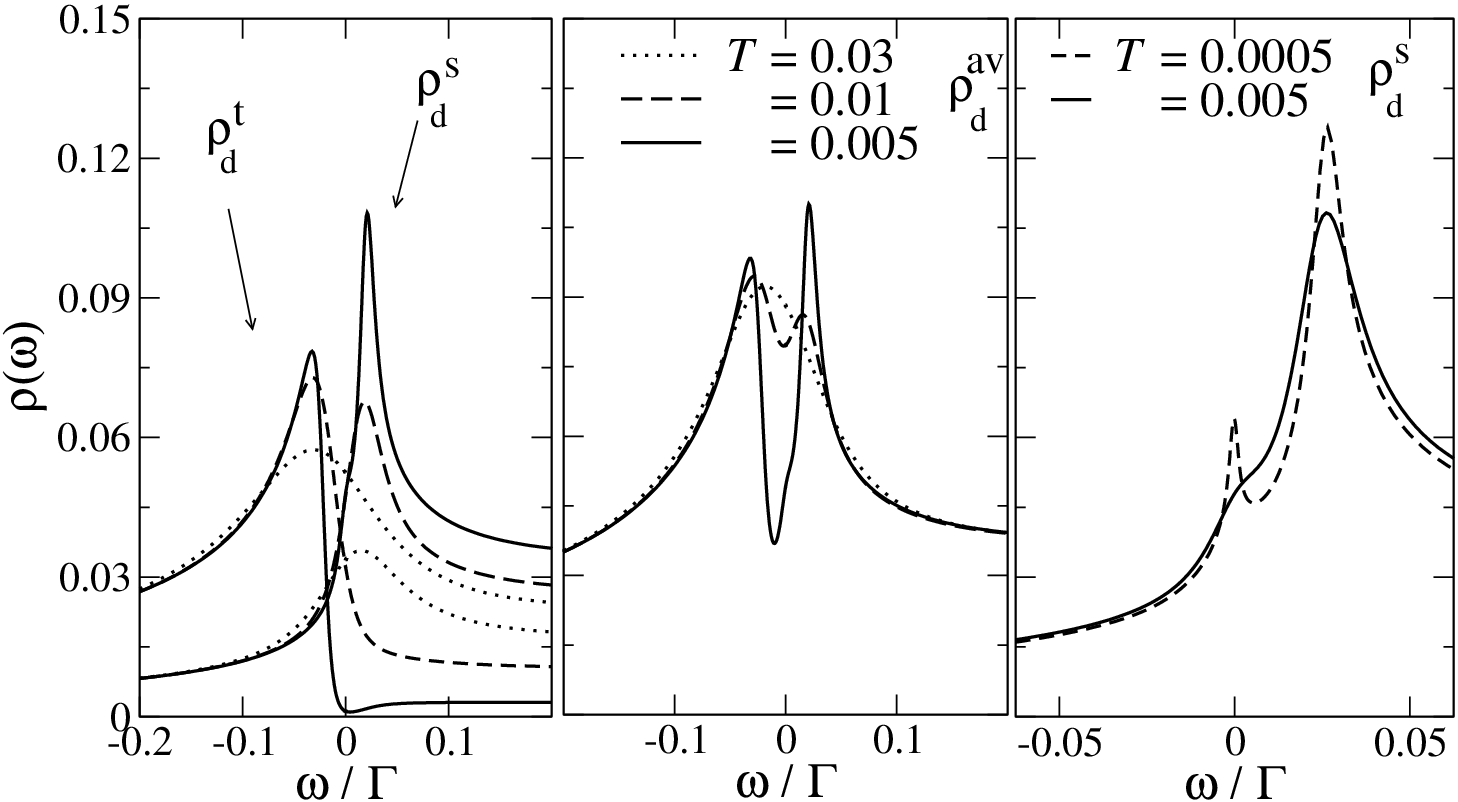}
\caption{Spectral densities as a function of frequency for $V=0$,  
$E_t=-2$, $E_s=-2.03$ and several temperatures. 
Left: $\rho_{d}^{s}(\omega )$ and $\rho_{d}^{t}(\omega )$, middle: weighted average [see Eq. (\ref{rhoav})],
right: $\rho_{d}^{s}(\omega )$ at low temperatures.}
\label{dsin}
\end{figure}

How do the spectral densities evolve as one
moves from the quantum critical line? 
For $V=0$, this is shown in Fig. \ref{dsin} on the singlet side of the transition.
For small temperatures, decreasing $E_{s}$ from the quantum critical line ($E_t=E_s$), 
$\rho_{d}^{s}(\omega )$ displaces to positive frequencies, while $\rho_{d}^{t}(\omega )$ 
decreases and displaces its weight to negative 
frequencies. As a consequence, a 
pseudogap opens in $\rho _{d}^\text{av}(\omega )$.
A similar pseudogap was found before in studies of two-level systems and interpreted 
as the low temperature part of a two-stage Kondo effect  \cite{hof}, along the
lines discussed in Sections \ref{even} and \ref{split}.
At very low temperatures (below $0.005 \Gamma$), a spurious spike 
appears at the Fermi energy
in $\rho_{d}^{s}(\omega )$. This is due to a known shortcoming of the NCA that takes place 
when the ground state for zero hybridization is non-degenerate, for example
under an applied magnetic field \cite{win}. However, as argued in Ref.  \cite{win}
it is interesting to note that this shortcoming does not affect the calculation
of thermodynamic properties under a finite applied magnetic field \cite{vil}.

\begin{figure}[tbp]
\includegraphics[width=12cm]{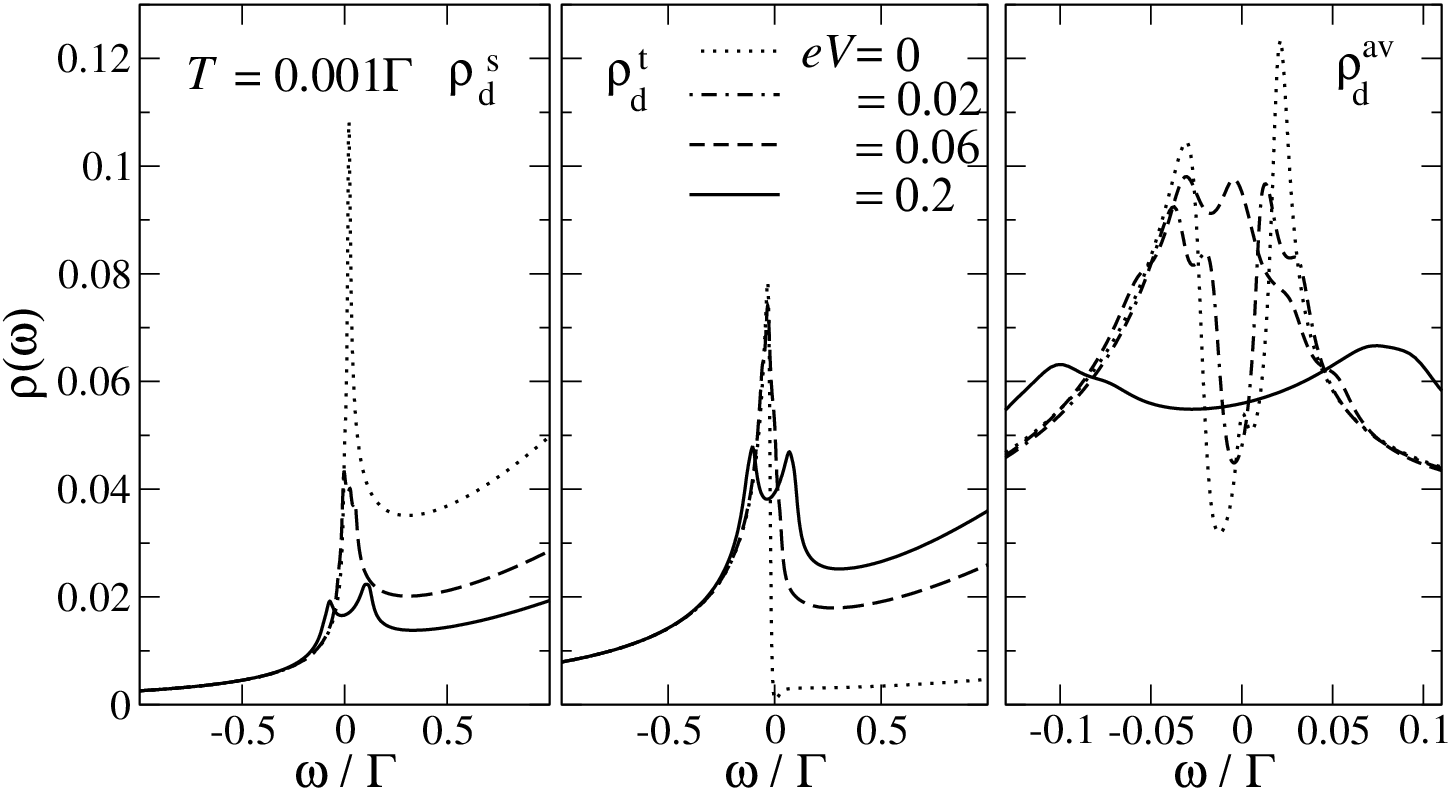}
\caption{Spectral densities $\rho_{d}^{s}(\omega )$ and $\rho_{d}^{t}(\omega )$ 
as a function of frequency for $T=0.001\Gamma$,  
$E_t=-2$, $E_s=-2.03$ and several bias voltages.}
\label{dsinv}
\end{figure}

The densities under an applied voltage are shown in Fig. \ref{dsinv}.
The main effect of the bias voltage $V$ is to lead to a decrease of the singlet
part of the density $\rho_{d}^{s}(\omega )$ and a simultaneous increase of the 
triplet part $\rho_{d}^{t}(\omega )$ for positive frequencies. In addition, near
the average Fermi level $(\mu_L + \mu_R)/2$, as  $V$  is increased, 
the pseudogap in the average density of states $\rho _{d}^\text{av}(\omega )$ first
closes, leading roughly to a single broader peak near $(\mu_L + \mu_R)/2$, and then, for
larger  $V$, this peak in $\rho _{d}^\text{av}(\omega )$ splits in two near
$\mu_L$ and $\mu_R$ as in Fig. \ref{doam} and the ordinary Anderson model \cite{win}.
The right panel of Fig. \ref{dsinv} shows $\rho _{d}^\text{av}(\omega )$ 
with more detail at low energies and including another small voltage. 
This fine structure suggests that as the gate voltage is applied, the peaks of
$\rho_{d}^{\eta}(\omega )$ split in two, shifted in $\pm eV/2$. Then, naturally,
the pseudogap closes when $eV$ reaches the difference between the position of the 
peaks, which is near $E_t-E_s$.

The spurious peak at the Fermi level disappears already for very small bias voltages.

\begin{figure}[tbp]
\includegraphics[width=8cm]{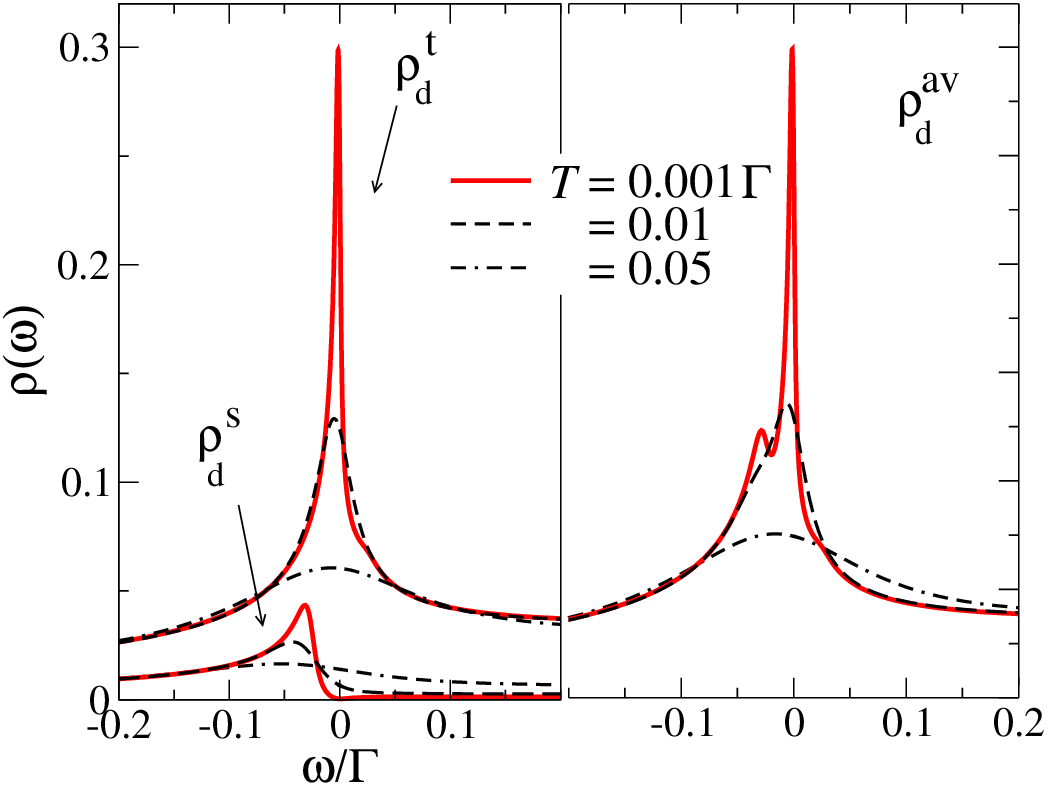}
\caption{Spectral densities $\rho_{d}^{s}(\omega )$ and $\rho_{d}^{t}(\omega )$ 
as a function of frequency for $V=0$,  
$E_t=-2$, $E_s=-1.97$ and several temperatures.}
\label{dtri}
\end{figure}

The spectral densities on the triplet side of the transition ($E_{s}>E_{t}$)
at equilibrium are shown in Fig. \ref{dtri}. 
In this case, the ground state for $V_s=V_t=0$ is degenerate and no spurious peaks appear.
Therefore our results are more robust. 
In contrast to the previous case, 
$\rho_{d}^{t}(\omega )$ remains peaked at the Fermi energy for low temperatures.
This is a consequence of the partial Kondo effect, by which the spin 1 at
the dot forms a ground state doublet with the conduction electrons of both
leads, as it is known from the exact solution of the model when the
singlet can be neglected \cite{gen,be1,be2} or the spin 1 underscreened Kondo model
\cite{s1,meh}.

The singlet part of the density $\rho _{d}^{s}(\omega )$ displaces to 
negative frequencies in this case. Therefore a 
pseudogap also appears in 
$\rho _{d}^\text{av}(\omega )$, but at finite
frequencies. Note that at high temperatures both densities  
are similar except for a constant factor and the difference in the structure 
of $\rho _{d}^{s}(\omega )$ 
and $\rho _{d}^{t}(\omega )$ develops at a characteristic temperature of the order
of a fraction of $|E_{s}-E_{t}|$.

\begin{figure}[tbp]
\includegraphics[width=8cm]{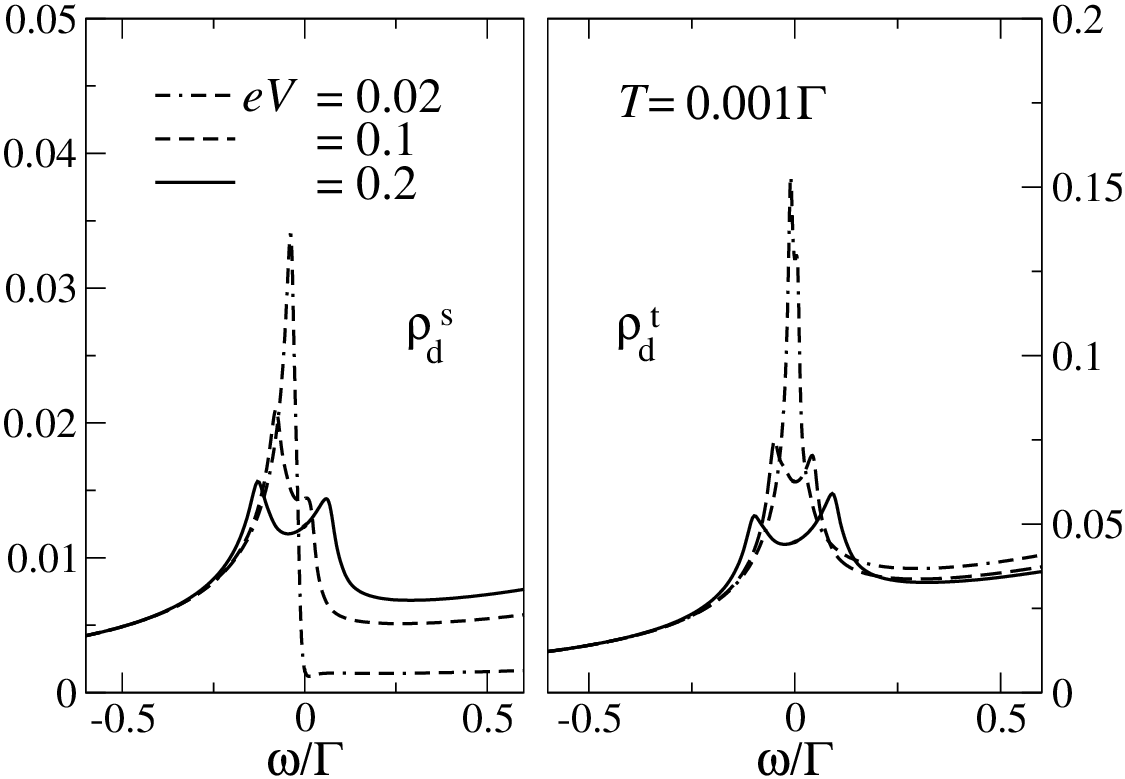}
\caption{Spectral densities $\rho_{d}^{s}(\omega )$ and $\rho_{d}^{t}(\omega )$ 
as a function of frequency $T=0.001\Gamma$,  
$E_t=-2$, $E_s=-1.97$ and several bias voltages.}
\label{dtriv}
\end{figure}

The effect of an applied bias voltage on these densities is shown  in Fig. \ref{dtriv}.
In this case, the singlet part of the density increases with voltage at positive
frequencies, in contrast to the case shown above (Fig. \ref{dsinv}) for the singlet side
of the transition. However, the behavior of $\rho _{d}^\text{av}(\omega )$ near
the average Fermi energy is similar as in the above case. The peaks at equilibrium
first broaden and merge into one for small bias voltage $V$, and for larger $V$, this
peak splits in two centered at energies near $\mu_L$ and $\mu_R$.

\subsection{The equilibrium conductance}\label{equi} 

\begin{figure}[tbp]
\includegraphics[width=8cm]{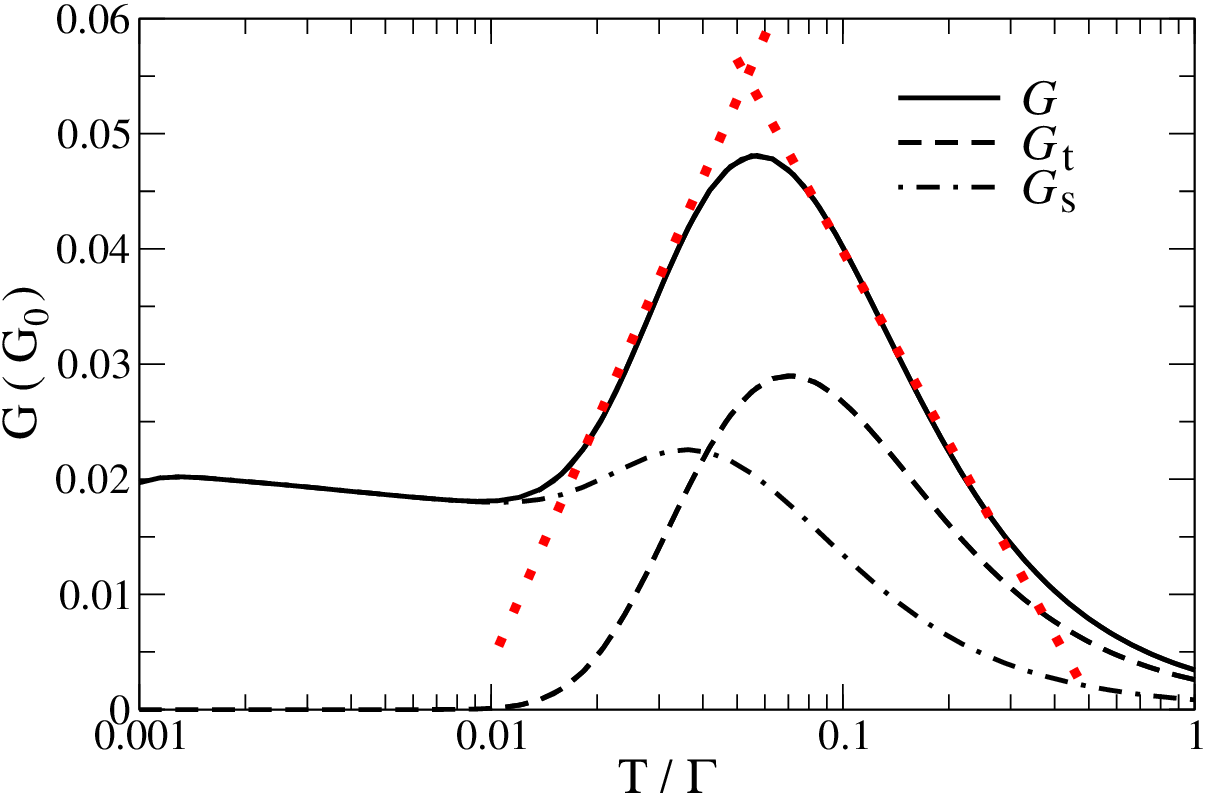}
\caption{Zero bias conductance $G(T,0)$ in units of $G_0=2e^2 A/h$ as a function of
temperature (full line) and contributions from the singlet 
(dashed dot line) and triplet 
(dashed line) for $E_t=-3, E_s=-3.1$.
Straight dot lines are guides to the eye.}
\label{gs}
\end{figure}

The conductance $G(T,V)=dI/dV$ for $V=0$ on the singlet side of the transition is shown 
in Fig. \ref{gs}.
together with the contributions of the singlet 
[$\Gamma ^{s}\rho_{d}^{s}(\omega )$ in Eq. (\ref{i})] and triplet 
[$\Gamma ^{t}\rho_{d}^{t}(\omega )$] part of the
spectral densities. Our result agree with previous ones using NRG \cite{logan,hof}
and with experiment \cite{roch}. In particular, the increase and decrease of 
$G(T)$ from its maximum value are logarithmic to a good degree of accuracy.
At very low temperatures (below 0.01 $\Gamma$ in Fig. \ref{gs}), 
our result for $G(T)$ increases slightly as the temperature is lowered,
while one expects a saturation at a value given by the generalized 
Friedel-Luttinger sum rule \cite{logan} (see below).
This low-temperature increase is due to the spurious peak that develops 
in $\rho_{d}^{s}(\omega )$
as a consequence of the NCA, as explained in Section \ref{spec}.

According to the generalized Friedel-Luttinger sum rule \cite{logan}, 
the conductance $G(T,V)$ at zero temperature and without applied bias voltage
on the singlet side of the transition
is given by
\be  
G_s(0,0)=G_0 \text{sin}^2 \left( \frac{\pi}{2} n_{\text{odd}} \right)  \label{g0s} 
\ee
where $G_0=2e^2 A/h$ and 
$n_{\text{odd}}=\langle \sum_{\sigma} |\sigma \rangle \langle \sigma| \rangle$
$=1-\langle |00 \rangle \langle 00|+\sum_{M} |1M \rangle \langle 1M| \rangle$
is the total occupation of the configuration with odd number of particles.
For the parameters of Fig. \ref{gs}, we obtain $n_{\text{odd}}=0.10 \approx 0$  and then
one expects a low value of $G_s(00)/G_0$. 
In fact inserting this value of $n_{\text{odd}}$ in Eq. (\ref{g0s}) one obtains 
$G_s(0,0)/G_0=0.0245$. Our corresponding result is near 0.020 (see Fig. \ref{gs}). 
Although it is known that the NCA does not satisfy Fermi liquid relationships, 
the deviation (near 0.005) is not too large for a magnitude that is in general of
the order of 1. For the particular case of Fig. \ref{gs}, the deviation is near 
10 \% of the maximum value shown in the figure.

\begin{figure}[tbp]
\vspace{1cm}
\includegraphics[width=8cm]{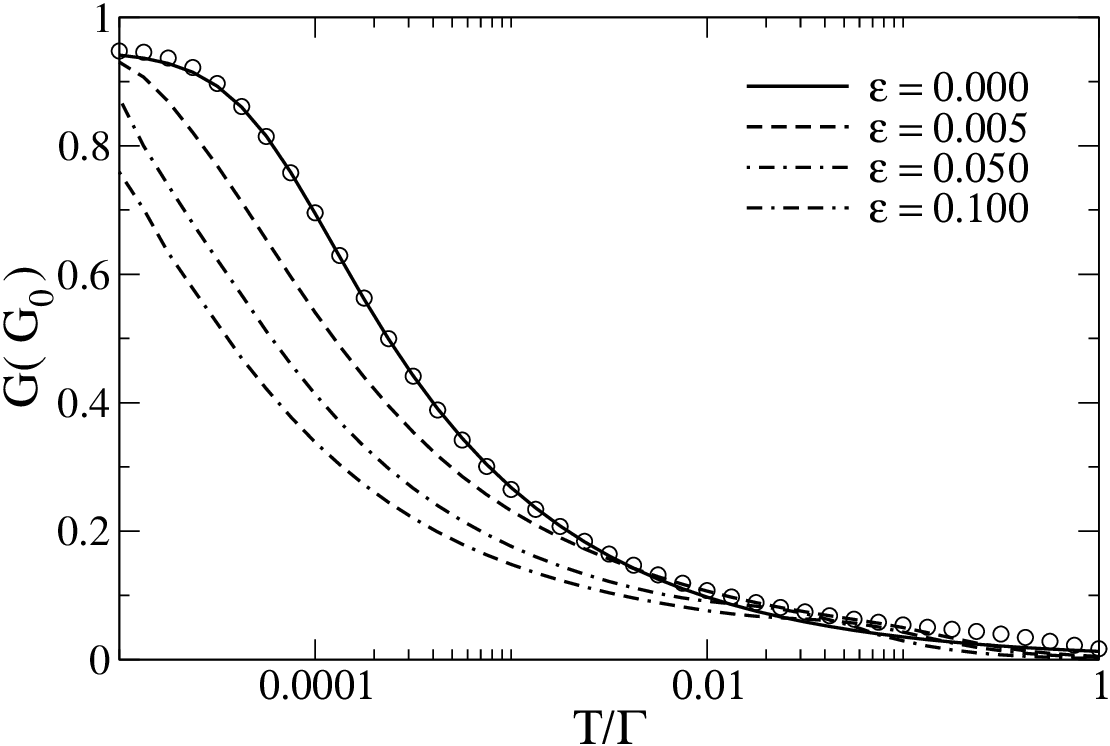}
\caption{Zero bias conductance $G(T)$ as a function of
temperature for $E_t=-3$ and 
several values of $\varepsilon= E_s-E_t$  
(increasing from top to bottom). 
The circles correspond to Eq. (\ref{ge}).}
\label{gte}
\end{figure}

The evolution of the zero-bias conductance as the system moves from the 
quantum critical line $E_t=E_s$ to the triplet side of the transition
$E_t<E_s$ is shown in Fig. \ref{gte}. At the transition, the conductance 
is the same as that of an OAM \cite{win}, with parameters given by the 
equivalence explained in Sections \ref{qcl} and \ref{dens}. In particular,
we find that the conductance is very well described by the empirical 
curve derived by fitting results of the NRG for a spin 1/2
\begin{equation}
G_{E}(T)=\frac{G(0)}{\left[ 1+(2^{1/s}-1)(T/T_{K})^{2}\right] ^{s}},
\label{ge}
\end{equation}
with $s=0.22$. As $E_t$ is lowered, removal of degeneracies in the
ground state leads to a decrease in the conductance at intermediate
temperatures. At zero temperature, the generalized Friedel-Luttinger 
sum rule \cite{logan}, for the triplet side of the transition gives
\be  
G_t(0,0)=G_0 \text{cos}^2 \left( \frac{\pi}{2} n_{\text{odd}} \right)  \label{g0t}.
\ee
Since the valence is only slightly increased by a small decrease in $E_t$,
one expects that $G(0,0) \approx G_0$ 
in good agreement with our
results. The temperatures reached in our NCA approach 
for the larger values of $\varepsilon= E_s-E_t$ used in Fig. \ref{gte} 
are not low enough to reach these high values of
the conductance. 

\begin{figure}[tbp]
\vspace{1cm}
\includegraphics[width=8cm]{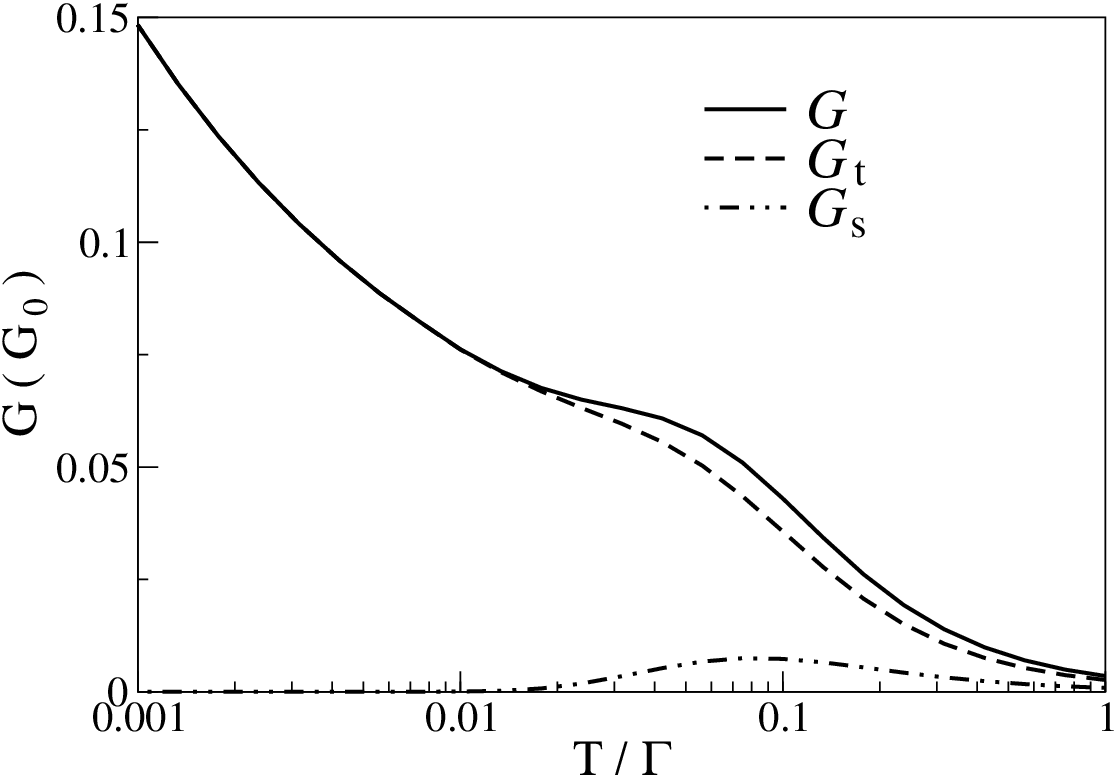}
\caption{Zero bias conductance $G(T)$ as a function of
temperature (full line) and contributions from the singlet (dashed dot dot line) and triplet 
(dashed line) for $E_t=-3, E_s=-2.9$.}
\label{gt}
\end{figure}

A distinct feature of $G(T,0)$ on the triplet side of the transition is
the developing of a bump or a plateau at intermediate temperatures.
This is more clearly displayed in Fig. \ref{gt} where the scale in the conductance 
has been expanded. This structure has not been noticed
in previous NRG calculations of the conductance in two-level systems \cite{hof,logan}.
This might be due to insufficient calculations in the appropriate range of 
temperatures (which correspond to rather small values of the conductance).
Another possible reason is the loss of resolution of NRG
for high-energy features. This shortcoming of NRG is clearly manifest \cite{va1,va2}
in systems of two QD's in which the Kondo resonance is split in two 
\cite{va1,va2,dias}. The separation of the spectral density $\rho _{d}^\text{av}(\omega )$
which enters the equation for the current (\ref{i}) into singlet and triplet components, 
as shown in Fig. \ref{gt}, shows that the bump is due to charge excitation involving 
the singlet component (which are peaked at an energy $E_s-E_t$, see Fig. \ref{dsin})
broadened by the temperature. 

This provides an interpretation of the corresponding transport experiments through
C$_{60}$ QD's (Fig. 4 (b) of Ref. \cite{roch}). The comparison suggests
that in general, the temperature in the experiment could not be lowered significantly 
after the plateau has been completed,
and that the conductance should continue to increase for decreasing temperatures. 
Roch \textit{et al.} \cite{roch}
suggested a different physical picture, fitting the plateau with the empirical
Eq. (\ref{ge}) with a smaller value of $G_0$ and speculated that the further increase
in $G(T,0)$ at smaller temperatures might be due to the opening another parallel transport 
mode. The effect of a second screening 
is expected to lead to a decrease of the conductance
on general physical grounds \cite{pus}. Our results also indicate that while
Eq. (\ref{ge}) is a good curve fitting for the conductance of the OAM, 
it does not work in the STAM for general 
values of the parameters. An exception is of course, the exactly solvable quantum critical 
line in which the STAM is mapped onto an OAM plus a free spin \cite{allub}, as
described in Section \ref{exact}.

\subsection{Conductance as a function of bias voltage}\label{nonequi} 

\begin{figure}[tbp]
\vspace{1cm}
\includegraphics[width=8cm]{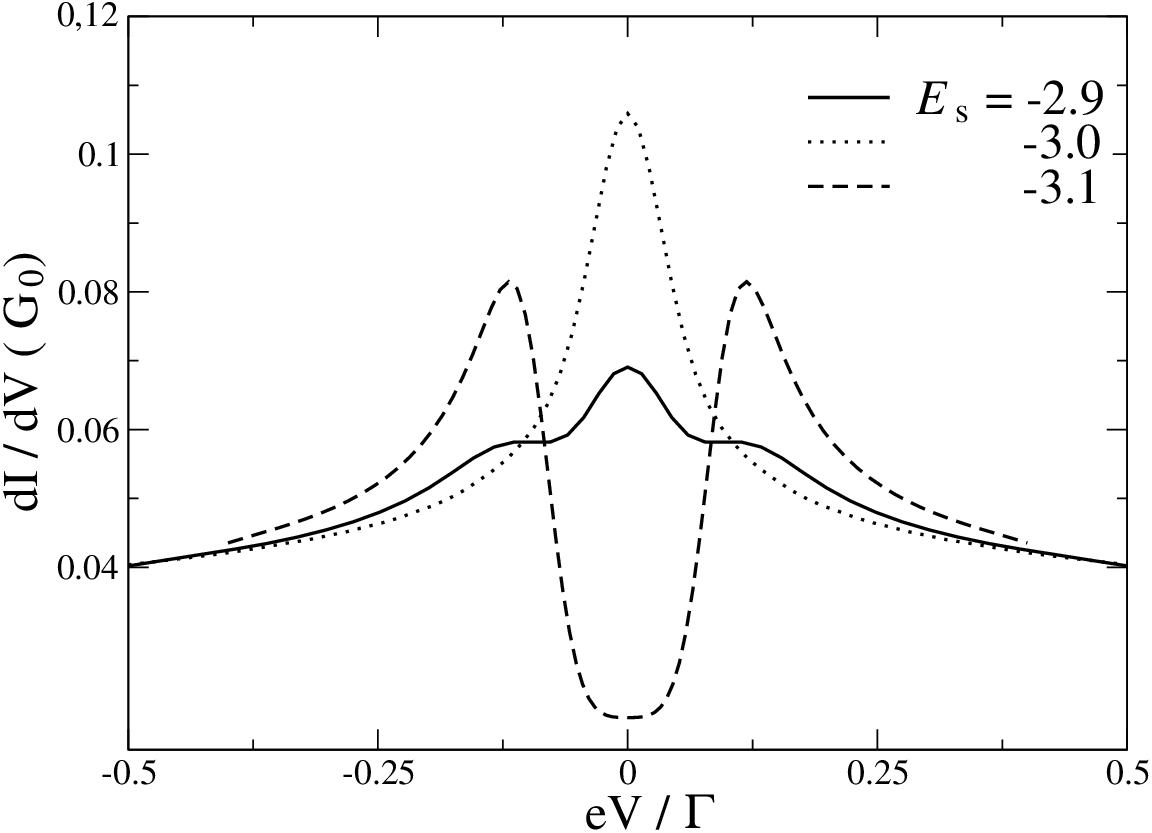}
\caption{Differential conductance as a function of
bias voltage $V$ for $T=0.01$, $E_t=-3$ and several values of $E_s$.}
\label{didv}
\end{figure}

In Fig. \ref{didv}  we show the differential conductance $G(T,V)=dI/dV$ as a function of bias voltage
for small temperatures and for three values of $\varepsilon= E_s-E_t$. One of them ($\varepsilon=0$) 
corresponds to the quantum critical transition and for the other two, the system is either on
the singlet ($\varepsilon <0$) or triplet ($\varepsilon>0$) side of the transition. The
remarkable change of behavior at the transition is evident. As expected from the results of
Section  \ref{spec}, the opening of a gap in $\rho _{d}^\text{av}(\omega )$ near the Fermi energy
on the singlet side of the transition, leads to a dip in $G(0,V)$ at low $V$. The width of this dip
is of the order of $|\varepsilon|$. At the bottom of the dip, the small value of the conductance $G(0,0)$ 
should be given by the generalized Friedel-Luttinger sum rule Eq. (\ref{g0s}),
while the NCA results have a deviation of nearly 20 \% of this value,
as discussed in Section \ref{spec}. For larger values of $-\varepsilon$, the dip
becomes wider, and the result is similar to the conductance observed in finite chains of 
an even number of Mn atoms on CuN \cite{hir}.

\begin{figure}[tbp]
\includegraphics[width=8cm]{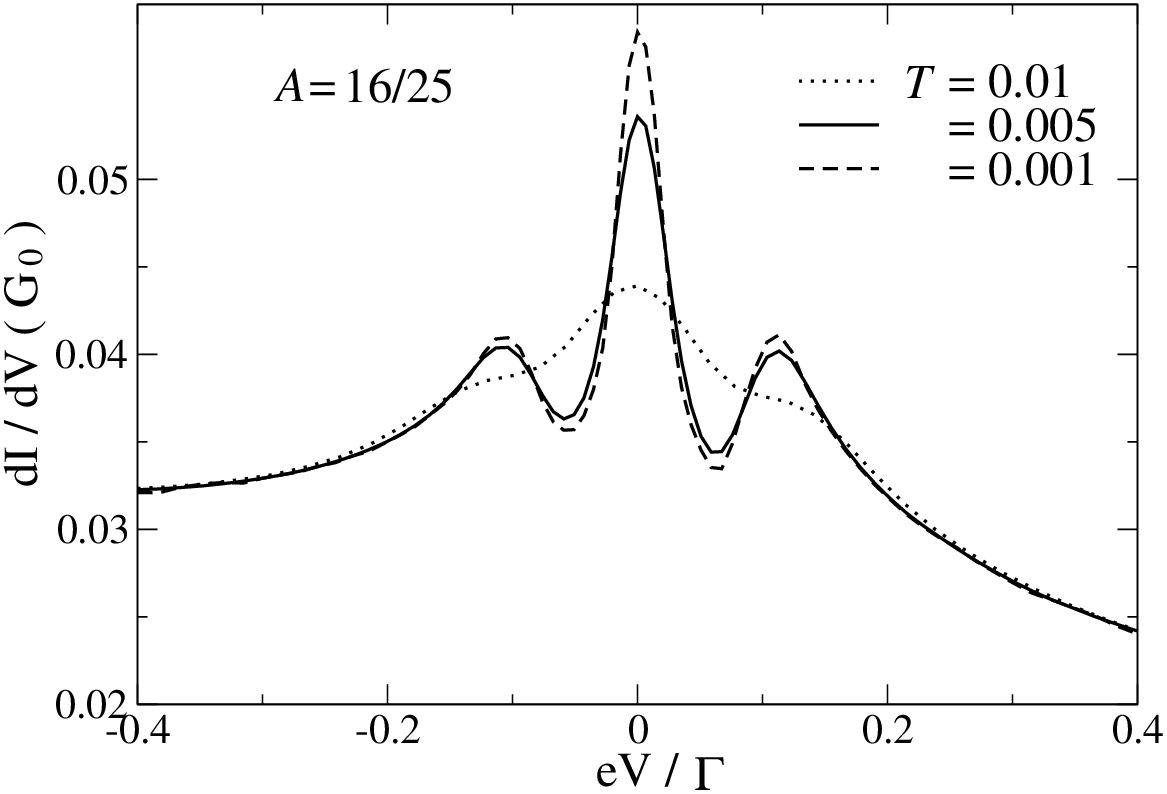}
\caption{Differential conductance as a function of
bias voltage for $E_t=-3, E_s=-2.9$ and several temperatures.}
\label{didvt}
\end{figure}

On the triplet side of the transition, a structure with a central peak at zero bias and
two lateral maxima is obtained. 
The three peaks are more marked at smaller temperatures,
as we have shown in Fig. 4 of Ref. \cite{ro0}. This structure also agrees
qualitatively with the experimental findings in C$_{60}$ QD's (Fig. 4 (a) of Ref. \cite{roch}). 
Actually the
observed structure is asymmetric in contrast to the results shown in Fig. \ref{didv}.
This is a consequence
of our assumption of a symmetric voltage drop. In fact the couplings $\Gamma _{L}^{\eta }$ 
and $\Gamma _{R}^{\eta }$ are usually very different for C$_{60}$ molecules \cite{sco} and
this leads to an asymmetric voltage drop. A reasonable assumption is that the voltage drop 
from one lead to the molecule is inversely proportional to the coupling to the corresponding
lead \cite{rpt}. 
In our calculation, the ratio between couplings enters through
the asymmetry parameter $A$ that controls the magnitude of the current 
[see Section \ref{cur} and Eqs. (\ref{i}) and (\ref{a})] and the functions
(\ref{avf}) that enter the selfconsistency equations (see Section \ref{nca}).
A calculation for an asymmetric voltage drop is shown in Fig. \ref{didvt}. 
We have taken $\mu_L=4eV/5$, $\mu_R=-eV/5$, and 
$\Gamma _{R}^{\eta }/\Gamma _{L}^{\eta }=4$ (assumed independent of $\eta =s$ or $t$), 
keeping the same sums $\Gamma _{R}^{\eta } + \Gamma _{L}^{\eta }$ as before. 
As expected, 
now the height of the lateral peaks is different, with the left peak as the second most
intense after the central one, in agreement with experiment. The nonmonotonic behavior 
of $G(0,V)$ can be again qualitatively understood from the structure of the spectral densities
discussed in Section \ref{spec}. Assuming as a first approximation that $\rho _{d}^\eta(\omega )$
does not depend on voltage, it is clear From Eq. (\ref{i}) that the conductance at
zero temperature $G(0,V)$ would proportional to the average of $\rho _{d}^\text{av}(\omega )$ 
in a window of $\omega $ of width $eV$ around the Fermi energy. Since $\rho _{d}^\text{av}(\omega )$
is peaked at the Fermi energy (as a consequence of the peak in the triplet part 
$\rho _{d}^t(\omega )$), $G(0,V)$ decreases with applied bias voltage $V$ for small $V$. However,
when the window of width $eV$ reaches the peak in the singlet contribution $\rho _{d}^s(\omega )$,
the average of the total spectral density  $\rho _{d}^\text{av}(\omega )$ is expected to increase 
and this leads to a peak in $G(0,V)$ at finite bias voltages. This explains the result
shown in Fig. \ref{didvt} at small temperatures. 

\begin{figure}[tbp]
\includegraphics[width=8cm]{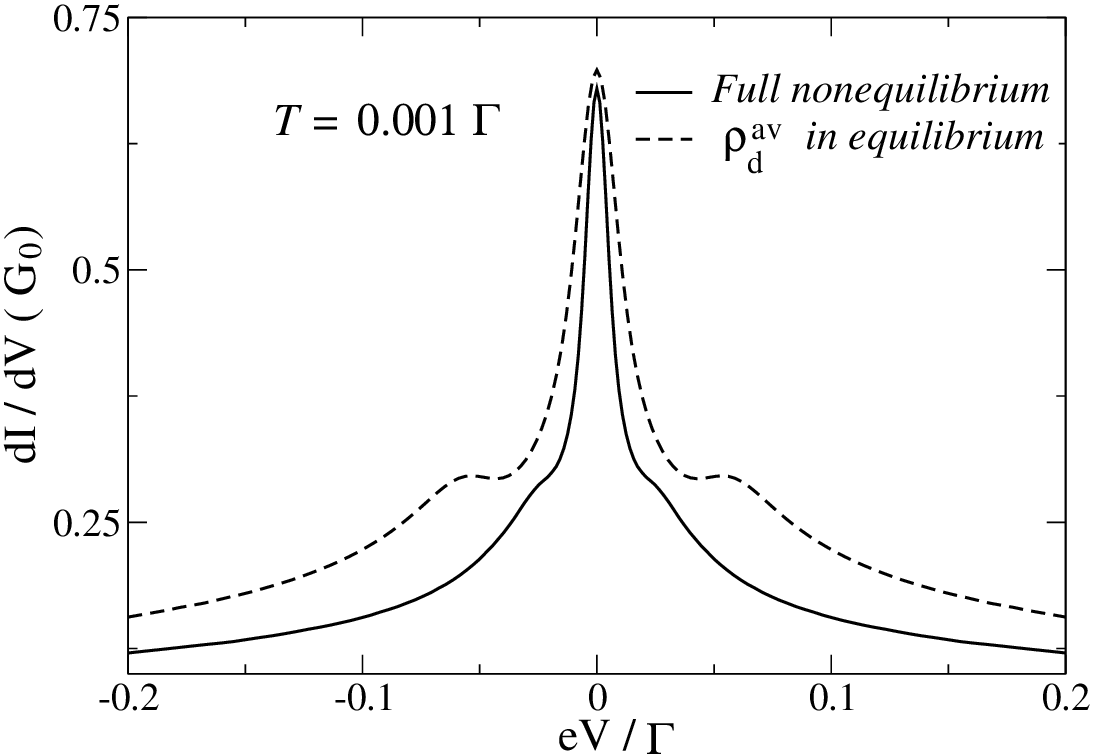}
\caption{
Full line: differential conductance as a function of
bias voltage for $E_t=-2, E_s=-1.97$ and $T=0.001\Gamma$.
Dashed line: corresponding result taking $\rho _{d}^\text{av}(\omega )$ at $V=0$.}
\label{didvt2}
\end{figure}

For other parameters, which would correspond to another experimental situation,
in particular nearer to the transition, 
it might happen that $\rho _{d}^s(\omega )$
broadens as a consequence of the applied bias voltage and the structure with three peaks is absent.
This is the case for the parameters of Fig. \ref{didvt2}, where instead of three peaks, 
we obtain two small shoulders at positive and negative voltages, as shown by the full line 
in Fig. \ref{didvt2}.
The three-peak structure is however the general behavior expected for a system well inside the Kondo
regime when $E_s-E_t$ is larger that the width of the peaks in the spectral densities.

Due to the difficulties in the calculation of nonequilibrium properties, an 
approximation usually made consists in taking the density of states calculated in equilibrium,
and assume that it is constant with applied bias voltage. 
This approach, valid for non-interacting systems has been used for example
in combination with first-principle calculations, to calculate the current
through single molecules \cite{heur}. Some results for the transport through 
two-level systems were obtained also in this approximation \cite{logan}, 
using a formula
equivalent to Eq. (\ref{i}), but with the average density calculated for $V=0$. 

From the result presented in Section \ref{spec} for finite $V$, it is clear that
this procedure is not valid in general. An example is shown by the dashed line
in Fig. \ref{didvt2}. When the equilibrium densities are taken instead of the 
nonequilibrium ones, the effects of broadening of the spectral densities caused
by the applied bias voltage are missed, and as a consequence, still the presence of two
side peaks (although broad) is predicted for the conductance, in contrast to the 
full nonequilibrium calculation.
Naturally, the splitting of the Kondo resonance at higher bias voltages is
also missed if frozen densities are assumed (see Section \ref{spec} and Ref \cite{win}).

\section{Summary and discussion}\label{sum}

We have studied the singlet-triplet Anderson model (STAM) in which a configuration with
a singlet and a triplet is mixed with another one with a doublet, via the hybridization
with a conduction band. This is the simplest model that describes the conductance
trough a multilevel quantum dot, hybridized with two leads in an effective one-channel fashion.
As found earlier in intermediate valence systems \cite{allub}, the model has a quantum
phase transition that separates a region with a singlet ground state from another one
with a doublet ground state. This transition has been studied recently in transport 
measurements through C$_{60}$ quantum dots  \cite{roch}. Our results provide an
explanation to the observed behavior at both sides of the transition. In particular,
in the region of parameters in which the ground state is a doublet, we obtain a zero bias 
conductance with a plateau at intermediate temperatures, which agrees with experiment. 
The three-peak structure
observed in the nonequilibrium conductance as a function of applied bias voltage is also 
explained by the model. The separation of the electronic spectral density at the dot 
into two parts, which correspond to excitations involving either the singlet or the
triplet, leads to a more transparent understanding of the underlying physics. In particular,
the above mentioned plateau and the observed peak at finite bias voltages are
due to singlet excitations.

We have also studied several limits of the model, which allow us to shed light on
the expected behavior of the conductance at very low temperatures and bias voltages ($G(0,0)$),
as the integer valence limits of the STAM, and an special quantum critical line in
which the model can be mapped into an ordinary Anderson model plus a free spin 1/2.
For $V_t=V_s$, the latter model with additional exchange interaction $H'$ [Eq. (\ref{oamd})]
is equivalent to the STAM. An analysis of $H'$ and the integer valent limits show that
the value of $G(0,0)$ is consistent with a generalized Friedel-Luttinger
sum rule derived recently \cite{logan}, and which takes into account that the system is 
a singular Fermi liquid when the ground state is a doublet, in a similar way as the
underscreened Kondo model  \cite{meh}. As it is clear form Eqs. (\ref{g0s}) and (\ref{g0t})
this means that an abrupt conductance change takes place at the transition \cite{logan}.
Although the NCA cannot reach zero temperature, our numerical results are consistent with
this result.
On the quantum critical line, the conductance is smoothly connected to that of the phase
with a doublet ground state.

Our NCA approach has the advantage over NRG that it can be rather easily extended to the 
nonequilibrium regime. When the configuration with an even number of particles is the favored one
(as we have assumed here), it works very well on the quantum critical line and 
one expects it to be accurate enough on 
the ``triplet'' side of the transition (where the ground state is a doublet). However,
on the  ``singlet'' side, a spurious peak in the singlet part of the spectral density
develops at very small temperatures and bias voltages, rendering our results quantitatively
inaccurate for these parameters. We have found that similar difficulties arise when the 
configuration with an odd number of particles is favored, but now on the triplet side
of the transition, including the quantum critical line. In fact, the ordinary Anderson model
that comes out of the mapping described in Section \ref{qcl} has now an occupation 
near zero and develops a spurious peak at low temperatures and bias voltages. 

While our results on the triplet side of the transition agree with the experimental 
results of Roch {\it et al.}  \cite{roch}, and the interpretation of them is rather 
simple, we cannot totally rule out the possibility that some of them are due
to the NCA approximation, and that the physical explanation of the observed
phenomena is different.
For example, the plateau of the conductance on the triplet side of the transition 
has not been reported in previous NRG calculations \cite{hof,logan}. We believe 
that this might be due to the lack of resolution of the NRG at finite frequencies \cite{va1,va2} 
This might be improved averaging over different shifted logarithmic discretizations ($z$ averaging) \cite{oli,zit}
and using recent developments (the full density matrix NRG) \cite{pete,wei}.
In any case, taking into account the difficulties to extend the technique out of equilibrium \cite{and}, it
seems that a combination of both techniques 
(numerical renormalization group and non-crossing approximation) 
might be suitable to obtain further progress.

\section*{Acknowledgments}

One of us (AAA) is supported by CONICET. This work was done in the framework
of projects PIP 5254 and PIP 6016 of CONICET, and PICT 2006/483 and 33304 of
the ANPCyT.


\begin{thebibliography}{99}
\bibitem{note2} Note that the slave-boson mean-field theory is an approximation
for the low-energy physics and is therefore inadequate to calculate
the Luttinger integral $I_L$ considered in Ref. \cite{logan}, which involves
high energies.
\end{thebibliography}
\end{document}